\documentclass[twocolumn,showpacs,nofootinbib,preprintnumbers,amsmath,amssymb,superscriptaddress]{revtex4}
\usepackage{epsfig}
\pdfoutput=1

\usepackage{amsmath,amssymb,amsthm,amsfonts}

\usepackage{ mathrsfs }
\usepackage{slashed}
\usepackage{graphicx}
\usepackage{subfigure}
\usepackage{color,rotating}
\usepackage{dsfont}
\usepackage{setspace}
\usepackage{verbatim}
\usepackage{fancyhdr}
\usepackage{hyperref}

\newcommand{\bea}{\begin{eqnarray}}
\newcommand{\eea}{\end{eqnarray}}

\def\eq#1{(\ref{eq:#1})}
\def\Eq#1{Eq.~(\ref{eq:#1})}
\newcommand {\apgt} {\ {\raise-.5ex\hbox{$\buildrel>\over\sim$}}\ }
\newcommand {\aplt} {\ {\raise-.5ex\hbox{$\buildrel<\over\sim$}}\ } 

\def\id{1\!\mbox{l}}
\def\s0#1#2{\mbox{\small{$ \frac{#1}{#2} $}}}
\def\0#1#2{\frac{#1}{#2}}

\def\CP{{\mathcal P}}

\def\CS{{\mathcal S}}

\def\CW{{\mathcal W}}

\def\CZ{{\mathcal Z}}

\newcommand{\tr}{\mathrm{tr}}

\newcommand{\be}{\begin{eqnarray}}
\newcommand{\ee}{\end{eqnarray}}

\def\empile#1\over#2{\mathrel{\mathop{\kern 0pt#1}\limits_{#2}}}

\def\beq{\begin{equation}}
\def\eeq{\end{equation}}
\def\bea{\begin{eqnarray}}
\def\eea{\end{eqnarray}}

\def\p{{\boldsymbol p}}

\def\d3p{\frac{d^3\p}{(2\pi)^3}E_\p}

\usepackage{color}

\newcommand{\Fig}[1]{Fig.~\ref{fig:#1}}
\newcommand{\fig}[1]{\ref{fig:#1}}

\newcommand{\Sect}[1]{Sect.~\ref{sec:#1}}

\renewcommand{\vec}[1]{\mathbf{#1}}

\catcode`\@=11

\newcount\@tempcntc
\def\@citex[#1]#2{\if@filesw\immediate\write\@auxout{\string\citation{#2}}\fi
  \@tempcnta\z@\@tempcntb\m@ne\def\@citea{}\@cite{%
        \@for\@citeb:=#2\do%
    {\@ifundefined{b@\@citeb}%
        {\@citeo\@tempcntb\m@ne\@citea%
                \def\@citea{,\penalty\@m\ }{\bf ?}\@warning%
                {Citation `\@citeb' on page \thepage \space undefined}}%
        {\setbox\z@\hbox{\global\@tempcntc0\csname b@\@citeb\endcsname\relax}
     \ifnum\@tempcntc=\z@ \@citeo\@tempcntb\m@ne%
       \@citea\def\@citea{,\penalty\@m}%
       \hbox{\csname b@\@citeb\endcsname}%
     \else%
      \advance\@tempcntb\@ne%
      \ifnum\@tempcntb=\@tempcntc%
      \else\advance\@tempcntb\m@ne\@citeo%
      \@tempcnta\@tempcntc\@tempcntb\@tempcntc\fi\fi}}\@citeo}{#1}}%

\def\@citeo{\ifnum\@tempcnta>\@tempcntb\else\@citea
  \def\@citea{,\penalty\@m}%
  \ifnum\@tempcnta=\@tempcntb\the\@tempcnta\else
   {\advance\@tempcnta\@ne\ifnum\@tempcnta=\@tempcntb \else
\def\@citea{--}\fi
    \advance\@tempcnta\m@ne\the\@tempcnta\@citea\the\@tempcntb}\fi\fi}

\catcode`\@=12
 
\begin{document}

\title{\bf Gauge turbulence, topological defect dynamics, and condensation in Higgs models}
 
\vspace{1.5 true cm}
 
\author{Thomas~Gasenzer}
\affiliation{Institut f\"ur Theoretische
  Physik, Universit\"at Heidelberg, Philosophenweg 16, 69120
  Heidelberg, Germany} \affiliation{ExtreMe Matter Institute EMMI, GSI,
  Planckstra{\ss}e 1, D-64291 Darmstadt, Germany} 

\author{Larry McLerran}
\affiliation{Physics Department, Bldg. 510A, Brookhaven National Laboratory,
   Upton, NY 11973, USA} 
\affiliation{RIKEN BNL Research Center, Bldg. 510A, Brookhaven National Laboratory,
   Upton, NY 11973, USA} 
\affiliation{Physics Department, China Central Normal University, Wuhan, China} 

\author{Jan M.~Pawlowski}
\affiliation{Institut f\"ur Theoretische
  Physik, Universit\"at Heidelberg, Philosophenweg 16, 69120
  Heidelberg, Germany} \affiliation{ExtreMe Matter Institute EMMI, GSI,
  Planckstra{\ss}e 1, D-64291 Darmstadt, Germany} 

\author{D\'enes Sexty}
\affiliation{Institut f\"ur Theoretische
  Physik, Universit\"at Heidelberg, Philosophenweg 16, 69120
  Heidelberg, Germany} \affiliation{ExtreMe Matter Institute EMMI, GSI,
  Planckstra{\ss}e 1, D-64291 Darmstadt, Germany} 

\begin{abstract}
  {The real-time dynamics of topological defects
   and turbulent configurations of gauge fields for electric and magnetic confinement
   are studied numerically within a 2+1D Abelian Higgs model.
   It is shown that confinement is appearing in such systems equilibrating after a strong initial quench 
   such as the overpopulation of the infrared modes.
   While the final equilibrium state does not support confinement, metastable vortex
   defect configurations appear in the gauge field which are found to be closely related to the appearance of
   physically observable confined electric and magnetic charges.
   These phenomena are seen to be intimately related to the approach of a non-thermal fixed point 
   of the far-from-equilibrium dynamical evolution, signalled by
   universal scaling in the gauge-invariant correlation function of the Higgs field.
   Even when the parametersÊof the HiggsÊaction do not support Êcondensate formationÊin 
   the	vacuum,	during this approach, transient Higgs condensation is observed.
   We discuss implications of these results for the far-from-equilibrium
   dynamics of Yang-Mills fields and
    potential mechanisms how confinement and condensation in non-abelian gauge fields 
    can be understood in terms of the dynamics of Higgs models.
    These suggest that there is an interesting new class of dynamics of strong coherent turbulent gauge fields with condensates.
}
\end{abstract}


\maketitle

\section{Introduction}\label{sec:introduction}

Matter produced in heavy-ion collisions has been argued to form a
\emph{Glasma} at early times
\cite{Kovner:1995ts,Krasnitz:1998ns,Krasnitz:2002mn,Lappi:2003bi,Lappi:2006fp}.
This Glasma is initially a highly coherent, stochastic ensemble of
colour electric and colour magnetic fields.  The colour fields are very
strong and correspond to high occupancy of coherent gluon modes.  The
gluon states begin evolving from certain initial conditions, but soon
fluctuations, amplified by instabilities, begin to dominate the form
of the classical fields.  The field configuration becomes turbulent,
but at the same time remains highly coherent, with a high occupancy of
gluon modes.

Initially, one particular scale characterizes the momentum
distribution of gluons.  This is the saturation momentum
$Q_\mathrm{sat}\gg\Lambda_\mathrm{QCD}$ which is derived from the
colour glass condensates of the initial nuclei
\cite{Gribov:1984tu,Mueller:1985wy,McLerran:1993ni,McLerran:1993ka}
As the system evolves, there is a separation of momentum scales
between an infrared scale $\Lambda_s$ and an ultraviolet scale
$\Lambda$, see e.g.\ 
\cite{Blaizot:2011xf,Kurkela:2011ti,Berges:2012ev,Berges:2013eia}.
The infrared scale $\Lambda_s$ marks the momentum below which modes
are maximally coherent and occupation numbers are of order $n(k) \sim
\alpha_s (Q_\mathrm{sat})^{-1}$, $k\lesssim Q_\mathrm{sat}$.  The
coupling constant of QCD is weak at that scale,
$\alpha_s(Q_\mathrm{sat} )\ll1$, since the saturation momentum is
large compared to the QCD scale for high energy collisions of large
nuclei.  Above the ultraviolet scale $\Lambda$, mode occupation
numbers go rapidly to zero.  At the initial time, the different scales
are of the same order,
\begin{equation}
  \Lambda_s(t_\mathrm{in}) = \Lambda(t_\mathrm{in}) = Q_\mathrm{sat}.
 \end{equation}
When thermalisation is reached at time $t_\mathrm{th}$, one has 
\begin{equation}
    \Lambda_s(t_\mathrm{th}) \sim \alpha_s \Lambda(t_\mathrm{th}). 
    \label{eq:therm}
\end{equation} 
The last relation follows because for a thermalized system, the
coherence scale is given by the magnetic mass $M_\mathrm{mag} \sim
\alpha_s T$.  Hence, relation \eq{therm} determines the thermalisation
time.

Studying these types of systems and initial conditions is interesting
in its own right, as they are relevant beyond the context of heavy-ion
collisions.  A central feature is that the considered systems
initially are dominated by modes with high occupation number, that is,
each of the momentum modes with momentum below the scale $\Lambda$ has
an occupation number much larger than one.  This is equivalent to
having large coherent fields.  If such a system were not described by
a gauge theory, one would expect the formation of a condensate
\cite{Svistunov1991a,Berloff2002a,Berges:2012us,Nowak:2012gd}.  
For gauge systems, however, it is
to a certain extent difficult to precisely articulate what is meant by
condensation, and one of the goals of the present article is to
demonstrate how this can be realized far from equilibrium.

It is well-known that in a generic evolution of a system of interacting 
fields towards a thermal state, turbulence can appear 
\cite{Frisch1995a,Zakharov1992a,Nazarenko2011a}.  
Also this is different for
gauge systems 
\cite{Berges:2008mr,Fukushima:2011nq,Berges:2012ev,Schlichting:2012es,
Kurkela:2012hp,Berges:2013eia,Fukushima:2013dma} 
as opposed to most other systems since the turbulent
variables describe very strong fields.  Thus, even though the coupling
can be very weak, one has both strong fields and strongly fluctuating
fields.  Such a system looks somewhat like the Yang-Mills vacuum,
where the strong quantum fluctuating fields can induce non-trivial
phenomena such as confinement.  One needs to ask here what the
expected non-trivial phenomena associated with such turbulent,
strong-field, but weakly coupled systems are.

In this article we discuss condensation as well as topological defects
and turbulence in abelian as well as non-abelian gauge theories.  We
present a concrete numerical study of the relation between turbulence,
topological defects, and condensation in an abelian Higgs model.  The
simulated evolutions start from a metastable symmetric state, leading
to tachyonic evolution, as well as from the above described
overpopulated initial states.  Our results show the transient
formation of (quasi-)topological vortex defects.  Vortices formed
during the early evolution after the initial quench are stabilized by
magnetic fluxes in the gauge sector.  At later times, these
destabilize due to strong short-wavelength fluctuations in the gauge
and matter sectors.  For sufficiently weak gauge coupling we find
soliton-like defect formation, separating spatial domains of
opposite-sign homogeneous charge distributions and thus exhibiting a
type of electric confinement.  
Also these patterns are due to vortex and vortex-sheet defects in the 
gauge-invariant field correlations of the Higgs field.  
These defects mark the onset of turbulent cascades
\cite{Nowak:2010tm,Nowak:2011sk,Gasenzer:2011by,Schole:2012kt,Karl:2013mn}.
As indirect cascades they lead to
symmetry breaking and condensation of the Higgs field \cite{Berges:2012us,Nowak:2012gd}.  
The universal scaling in the turbulent occupation number distributions
signals the approach of a non-thermal fixed point \cite{Berges:2008wm}.
Dynamics near such fixed points has recently been studied for non-relativistic Bose systems
\cite{Gasenzer:2008zz,Scheppach:2009wu,Nowak:2010tm,Nowak:2011sk,Schole:2012kt,Schmidt:2012kw,Karl:2013mn},
pure scalar relativistic theories
\cite{Micha:2004bv,Berges:2008wm,Berges:2008sr,Berges:2010ez,Gasenzer:2011by,Berges:2012us}
as well as pure gauge systems
\cite{Berges:2008mr,Berges:2012ev,Schlichting:2012es,Kurkela:2012hp,Berges:2013eia}
studied extensively in the recent past.  Most remarkably, increasing
the gauge coupling causes the condensation of the Higgs field to
disappear.

Our article is organized as follows.  In \Sect{Condensation}, we
discuss generic issues related to condensation phenomena that are
peculiar to gauge theories, as well as possible observables for
condensation.  We furthermore discuss how topological objects such as
vortices and defects may appear in turbulent gauge systems.  In
\Sect{AbelianHiggs} we lay out our results of numerical semi-classical
simulations of the equilibration dynamics of the Abelian Higgs model
starting from different initial states far from equilibrium.  These
results show the formation of vortical topological defects, related to
magnetic as well as electric confinement phenomena, the approach of a
non-thermal fixed point, as well as transient Bose-Einstein
condensation.

\section{Condensation in Gauge Theories}
\label{sec:Condensation}
\label{sec:vortices+turbulence}

\subsection{Confinement in Higgs models}

For gauge theories, condensation is conventionally formalized within Higgs models.
For example, in the Abelian Higgs model, a gauged vector potential couples to
a scalar field. 
The abelian $U(1)$-symmetric model is described by its classical action
\begin{eqnarray}
\nonumber 
  S[A_\mu,\phi]
  &=& -\int_x \left[ \014 F_{\mu\nu} F^{\mu\nu}
  +(D_\mu \phi)^* D^\mu \phi +V(\phi)\right],
\end{eqnarray}
%
%
\begin{eqnarray}
  V(\phi)
  &=& {1\over 2 }m^2 |\phi|^2 + 
{\lambda\over 4! } |\phi|^4 +  {3 m^4 \over 2 \lambda }  =   \0{\lambda}{4!}( \phi^*\phi-v^2)^2,
\nonumber\\
\label{eq:U1HiggsModel}
\end{eqnarray}
where $\int_x = \int d^{1+2}x$, $D^{\mu}=\partial^{\mu}+ieA^\mu$, and 
$v= |m| \sqrt{6 / \lambda} $ is the Higgs field's equilibrium expectation value. 

The scalar field is in the fundamental representation of the gauge group, i.e., it acquires a phase under gauge transformations. 
 In the most basic setup, one considers the Higgs
potential and looks for a minimum at a non-zero value of the scalar
field.  The scalar field, however, rotates under local gauge
transformations and hence cannot have a non-zero expectation value.
The root of this problem is that the Higgs field possesses two degrees
of freedom, its phase and its modulus, of which the phase is gauge
dependent.  The modulus is positive definite, so that when
fluctuations are included, it will always have an expectation value.
However, at first glance, there is no gauge invariant way of
characterizing the condensed phase in terms of spontaneous symmetry breaking.

The Abelian Higgs model, nevertheless, shows several phases.  
In the condensed phase, one observes the Meissner effect of magnetic
fields being expelled by the charged Higgs field which, thinking in the Landau-Ginzburg condensed matter context, describes the Cooper pair condensate.
Moreover, depending on the value of the Ginzburg-Landau parameter which measures the relative strength 
of the Higgs and the gauge couplings, the superconductor is in its type I or type II phase.
In the type II phase, magnetic fields exceeding a critical strength become confined within
flux tubes, centered at Abrikosov vortices in the Cooper pair condensate.
Despite the fact that the phase of the Higgs field is gauge dependent, it is twisted by a non-zero integer multiple of $2\pi$ when following it around a vortex core. 
Gauge transformations cannot unwind this defect structure.  
Finally, the Abelian Higgs model can also be in the symmetric or Coulomb phase where magnetic flux is not confined.

Little is known about the role of topological defects for Higgs models excited far from thermal equilibrium.
In \Sect{AbelianHiggs} we will study numerically time-evolutions of the Abelian Higgs model in view of the role of defects in the equilibration dynamics after a strong initial quench.
Before we proceed with this, we discuss, in the remainder of this section, potential implications for non-Abelian gauge theories, in particular the relation between condensation, confinement, and topological defects. 
We will discuss monopole parametrisation defects in the non-Abelian gauge field which are understood to be closely related to electric confinement.
Interestingly, similar, vortex-type defects will show up in the simulations of the Abelian Higgs model in \Sect{AbelianHiggs}.

In non-Abelian Higgs models, condensation is associated with an analogous but in general richer spectrum of topological defects.
A way to formalize this structure is in terms of the following two order parameters: 
One is the conventional Wilson loop measuring confinement of electric charges.  The other is
the 't Hooft loop that measures the confinement of magnetic monopoles.
The Wilson and 't Hooft loops are convenient parameters to use for
systems in thermal equilibrium, that can be formulated in terms of a
Euclidean path integral.  

Three possible phases can be associated with these order parameters in
gauge theories \cite{Greensite:2011zz}: 
\\
--- The \emph{confined phase} shows electric confinement and magnetic
deconfinement. The Yang-Mills theory vacuum and the strong-coupling limit of 
compact electrodynamics provide example realisations of the confined phase.
\\
--- The \emph{magnetic confinement} phase shows electric deconfinement
and magnetic confinement.  The Abelian Higgs model in the Higgs phase
is an example of the magnetic confinement phase.
\\
--- The \emph{Coulomb phase} shows electric and magnetic deconfinement.
Electrodynamics without condensation is an example of the Coulomb phase, as there is confinement neither in the electric nor in the magnetic sector.  
  
A non-trivial example of the Coulomb phase occurs for $SU(2)$
Yang-Mills theory with an adjoint-representation Higgs field 
as it is realized in the (Georgi-Glashow) non-Abelian Higgs model.  An
expectation value of the Higgs field causes two vector bosons to
become massive.  One direction remains unbroken, so that both colour
electric and colour magnetic charges are deconfined.

The dynamics of all of these models is entirely non-trivial in the
infrared.  To generate electric confinement, one presumably needs
condensation or degeneracy with some type of colour magnetic monopole
excitations.  In analogy to the non-Abelian Higgs model, 
colour electric charge condensation is expected
to be required for magnetic confinement.

\subsection{Topological defects, condensation, and confinement in Yang-Mills theories} 

From the above discussion of condensation and confinement in Higgs
models it remains unclear whether analogous relations exist in pure
Yang-Mills theories which contain only adjoint-representation fields.
It has already been pointed out that the condensation of
adjoint-representation fields as in the Georgi-Glashow model generates
a Coulomb phase.  While the confinement of magnetic or electric flux
can be fairly easily imagined if there is condensation of
fundamental-representation fields, it remains unclear how this is
realized in a model with adjoint-representation fields only.  In this
subsection we present a possible order parameter for the
confinement-deconfinement transition in the frame of a non-Abelian
Higgs reformulation of the Yang-Mills Lagrangian.

The main difficulty in describing condensation phenomena in  
gauge theories is to find suitable gauge-invariant order parameters.
Gluon descriptions in terms of the gauge fields
$A_\mu^a$ are not gauge invariant, and it may be difficult to
directly read off physical mechanisms of condensation from
correlation functions of the gauge field.  This problem already occurs
in a static setting and has been discussed at length in the context of
the confinement-deconfinement phase transition
\cite{Greensite:2011zz}.

Standard confinement scenarios are based on condensation or
percolation involving topological defects, i.e. colour-magnetic monopoles 
or center vortices.  In QCD, in four dimensions, these topological 
defects are not stable objects as the related topological invariant 
vanishes.  The only stable configurations known are instantons.  
In fact, gauge field configurations which contain monopoles or 
vortices have infinite action.  
Nevertheless, non-trivial vacuum configurations are possible which carry a
non-vanishing topological density well-described in terms of these
defects instead of instantons.  This favors a description of the
ground state in terms of defects which are simplified within
appropriate gauge fixings.

In order to describe the static confinement-deconfinement phase transition in this way,
 an appropriate gauge is the Polyakov gauge where the temporal gauge field
is locally rotated into the Cartan subalgebra and made static.  For the non-equilibrium
evolution we consider in the following we take a spatial component, say $A_3$, and
apply such a gauge fixing.  More precisely, we apply a diagonalisation
transformation to the Wilson loop in $x_3$-direction,
\begin{equation}\label{eq:Wilson} 
\CW_3 = \CP \exp \left\{ i\,g\,\int_0^{L_3} d x_3\, A_3(x)\right\}\,,
\end{equation} 
where $\CP$ denotes path ordering and $L_{3}$ is the spatial extent along
this direction.  Now we write the Wilson line in terms of an algebra-valued
field $\phi$, to wit
\begin{equation}\label{eq:Higgs} 
\CW_3 = \exp \{i\, \phi\}\,, 
\end{equation} 
where $\phi$ is referred to as a Higgs field.  Under gauge
transformations $U\in SU(N)$ the Wilson loop \eq{Wilson} transforms as
\begin{equation}\label{eq:gaugeU} 
\CW_3 \to  U^{-1}(0)\CW_3 U(L_3)\,,
\end{equation} 
where $U(0)$ and $U(L_3)$ is the gauge-group element evaluated at
$x_3=0$ and $x_3=L_3$, respectively.  The $SU(N)$-rotation in
\eq{gaugeU} can be used to diagonalise the Wilson loop, and hence
$\phi$, up to defects, see e.g.\ \cite{Ford:1998bt}. A diagonal
$\CW_3$ takes a particularly simple form in terms of $\phi$.  For
example, for $SU(2)$, its trace reads
\begin{equation}\label{eq:cos}
\012 \tr \CW_3=\cos(\varphi/2)\,,\qquad {\rm with} \qquad \phi=
\varphi \,\0{\sigma^3}{2}\,.
\end{equation}
The above diagonalisation can be achieved within a whole class of
gauges.
The natural one is a type of Polyakov gauge where the $3$-component
of the gauge field is rotated into the Cartan subalgebra,
\begin{equation}\label{eq:polyakov} 
  A_3(x)= A_3^c(\bar x)\tau^{c}\,,\quad {\rm with}\  
  \bar x= (x_0, x_1, x_2 , x_3=0)\,,
\end{equation} 
where $\{\tau^c\}$ are the generators in the Cartan, i.e.\ for $SU(2)$
we have one Cartan component with $\tau^c=\sigma^3/2$, see also
\eq{cos}. In the gauge \eq{polyakov}, we have the relation
\begin{equation}\label{eq:phiA}
\phi= { g\,L_3} A_3^c(\bar x)\tau^{c}\,.
\end{equation} 
The definition of $\phi$ implies that its eigenvalues $\varphi_n$ with
$n=1,..., N_c$ in the fundamental representation are gauge
invariant. They are directly related to the eigenvalues of the Wilson
line $\CW_3$ which read $\exp\{ i\varphi_n\}$ and do not change under
gauge rotations \eq{gaugeU} with periodicity $U(L_{3})=U(0)$. We
emphasise that $\CW_3$ is in general not gauge invariant.

 In $SU(2)$, a center flip combined with an adjungation of the
 Polyakov loop is provided by the transformation $\varphi\to 2\pi
 -\varphi$ \cite{Ford:1998bt}, with the fixed point $\varphi=\pi$.  We
 conclude that in the center-symmetric phase where the trace of the
 Polyakov loop vanishes, we have $\langle\varphi\rangle =\pi$.  In
 euclidean space, a vanishing ground-state expectation value of the
 trace of the Polyakov loop implies confinement.  Hence $\langle
 \phi\rangle$ or $\tr\CW_3( \langle \phi\rangle) $ serve as well as
 order parameters of confinement as does $\langle
 \tr\CW_3\rangle$. Indeed one can also show that
\begin{equation}\label{eq:Jensen}
\tr \CW_3( \langle \phi\rangle)\geq  \langle\tr \CW_3\rangle\,,
\end{equation} 
with saturation in the confined phase, see
\cite{Braun:2007bx,Marhauser:2008fz}. The related order-parameter
potential has been computed perturbatively,
\cite{Gross:1980br,Weiss:1980rj}, and non-perturbatively,
\cite{Braun:2007bx,Marhauser:2008fz,Fister:2013bh}, in Yang-Mills
theory, see also
\cite{Diakonov:2012dx,Greensite:2012dy,Langfeld:2013xbf} for recent
lattice computations.

As a result of the above gauge prescriptions one obtains the Higgs
field $\phi$ as an order parameter field for the
confinement-deconfinement phase transition.  The parametrisation of
this phase transition in terms of $\phi$ allows to relate confinement
to the vortex percolation picture: The gauge described above which
diagonalizes $\CW_{3}$ potentially has defects that are located at the
points where $\CW_3$ is an element of the center $\CZ$ of the gauge
group; $\CZ\simeq Z_N$ for $SU(N)$, i.e., when $\varphi\in\{0,2\pi\}$
for $SU(2)$.  This is easily seen for the case $\CW_3=\id$.  Assume
that the Higgs field vanishes, $\phi=0$.  This can happen either
homogeneously, or with a non-trivial angular dependence around the
point in the $(x_{1},x_{2})$-plane where $\phi=0$.  In this case the
phase
\begin{equation} 
\hat \phi=\frac{\phi}{\|\phi\|}\,. 
\end{equation}
possesses a non-vanishing winding. 
The related topological invariant is the standard Hopf winding number
\begin{equation}\label{eq:Hopf}
  n(\CS)=\frac{1}{16 \pi\,i}\oint_\CS d^2 x\,\epsilon_{ij} \tr\, \hat\phi 
\,\partial_i \hat\phi\,\partial_j\hat \phi\,, 
\end{equation}
$\epsilon_{ij}$ being the antisymmetric tensor. 
Hence, magnetic (anti-)monopole defects occur, where $\CW_3=\id$ ($\CW_3=-\id$).
If $\langle\tr \CW_3\rangle=0$ then monopoles and anti-monopoles are condensed in equal proportions, i.e., no net magnetic charge exists and the system is in the confined phase.

Note that the
winding numbers \eq{Hopf} are parametri\-sation-windings and not windings of
the Wilson line $\CW_3$.  However, they are related to the monopole
number of $2+1$-dimensional Yang-Mills theory.  In four-dimensional
Yang-Mills theory they are known to be related to the instanton number, see e.g.\
\cite{Ford:1998bt}.  For vortex-free configurations, that is, those
where $\phi$ sustains no Hopf windings, the diagonalisation can be
performed in the entire space.

Rewriting the
Yang-Mills action in terms of the scalar field $\phi$ in an expansion
in $\phi$ leads to a gauge action of the remaining spatial components of the
gauge field coupled to the scalar field $\phi$. 
In order to describe confinement, the effective
potential of this field must exhibit, in the confined phase, a non-trivial minimum 
at $\varphi = \pi$. 

The Yang-Mills action $S_{\rm YM}= 1/2 \int_x \tr F_{ \mu
  \nu}^2$ written in terms of a reduced gauge theory with
Lorentz indices $\bar \mu= 0, 1, ..., d-1$ and a Higgs field formed from the remaining gauge field, reads,
in the Polyakov gauge \eq{polyakov},
\begin{equation}
  S_{\rm YM}= \012 
  \int_x\, \tr  F_{\bar \mu\bar \nu}^2+ \int_x\,\tr (D_d A_{\bar \mu})^2 
  +\int_x\,\tr (\partial_{\bar \mu} A_{d})^2 \,, 
\label{eq:action}\end{equation}
where $\int_x = \int d^{1+d} x$. If we restrict ourselves to
configurations that do not depend on $x_d$ but only on $\bar x=(x_0,
x_1 ,..., x_{d-1})$, this action further reduces to the Glasma action
\begin{equation}
 \01{L_d} S_{\rm YM}= \012 
  \int_{\bar x} \tr  F_{\bar \mu\bar \nu}^2+ \int_{\bar x}
  \tr\, (D_{\bar \mu} A_{d})^2 \,.
\label{eq:action_reduced}\end{equation}
In summary the following picture emerges. Yang-Mills theory, if
parameterized in diagonalisation gauges, resembles a non-Abelian
Higgs model, with the Higgs field in the adjoint representation. 
The Higgs field carries 
information about a confinement-deconfinement phase transition.
The diagonalisation gauges feature topological configurations/defects
which, at face value, are nothing but parameterisation defects even
though the global information about these defects relates to the
stable topological charge in these systems. In Yang-Mills theory,
the single defects are not stable and can decay. Despite this fact, they
are still related to stable topological configurations in Yang-Mills
theory and can be used to extract the topological density.
Interestingly, similar, vortex-type defects will show up in the simulations of the Abelian Higgs model in \Sect{AbelianHiggs}.
 
This establishes a close link between the two classes of
theories, pure Yang-Mills and Higgs, and makes it even more relevant to study the
far-from-equilibrium dynamics of the Higgs model.

\section{Turbulent dynamics of the Abelian Higgs model}
\label{sec:AbelianHiggs}

\subsection{Model and observables}

In the following we present our results of real-time semi-classical
simulations of an abelian $U(1)$-symmetric Higgs model.  
We will study the dynamical equilibration of $2+1$-dimensional systems starting from different initial conditions far from thermal equilibrium and different values of the Landau-Ginzburg parameter. 
The model is
described by the classical action \eq{U1HiggsModel}.
The corresponding equations of motion (EOM) for the gauge and scalar fields read
\begin{equation}\nonumber 
  \partial_\mu F^{\mu\nu}=J^\nu\,,\qquad 
D_\mu D^\mu \phi = \0{\partial V}{\partial \phi^*}\,, 
 \label{eq:eoM}
\end{equation}
with the current 
\begin{equation}\label{eq:current}
  J^\nu = i\, e \left(\phi\partial^\nu\phi^*-\phi^*\partial^\nu\phi\right) 
+2 e^2 \phi^*\phi A^\nu\,.
\end{equation}
The action and the EOM are discretized on a cubic 
lattice using the compact formulation for $U(1)$ gauge fields,
\bea
S[U,\phi] 
&=& - { 1\over \gamma e^2 a_s } 
\sum_{ i<j,x} (1 - \textrm{Re}\, U_{ij}(x)  )  
 \nonumber
\eea
\bea
&&+\ { \gamma \over e^2 a_s } \sum_{i,x} (1 - 
\textrm{Re}\, U_{0i}(x) )   
\nonumber
\\
&&+\ a_s^2 a_t\sum_x  \left( |D_\mu \phi(x)|^2 + V(\phi(x)) \right),
\eea
in terms of the plaquette
variables $U_{\mu\nu}$, and the ratio of spatial and temporal lattice 
spacings $ \gamma = a_s/ a_t $. The variation of the action with respect to 
the fields yields the discretised EOM as well as the Gauss constraint. 
Rescaling the scalar field as $ \phi' = \sqrt{\lambda} \phi$, one finds that 
only the ratio of the couplings $ e^2 /\lambda$ is relevant 
for the dynamics. 
This is called the Ginzburg-Landau parameter
\bea
\xi = {6 e^2 \over \lambda }
\eea
which controls the transition between the type I ($\xi>1/2$) and type II ($\xi<1/2$) superconducting phases.

During the time evolution we measure photon number distributions using 
the two-point correlators of the gauge fields,
\bea
 f^E_{ij}(\vec x-\vec y,t) =\langle E_i(\vec x,t) E_j(\vec y,t) \rangle_\mathrm{cl},
\eea
where the brackets $\langle ... \rangle_\mathrm{cl} $ denote averages over the classical ensemble defined by averaging over the initial conditions.
The occupation number is calculated from the spatial Fourier transform of the two-point correlators,
\bea
 f_E(\vec k,t) = \int d^2 x\, e^{i\vec k\vec x} f^{E}_{ii}(\vec x,t) 
=  {1\over V} \langle |E_i(\vec k,t)|^2\rangle_\mathrm{cl}\,,
\eea
which can be conveniently calculated 
using the Fourier transform $E_i(\vec k,t) = \int d^{2}{x}\,
E_i(\vec x,t)\, e^{i\vec k\vec x} $ of the field variables. 
We consider a definition of the occupation number in terms of the field and its canonical conjugate which does not involve the dispersion relation explicitly.
The Coulomb gauge is used to evaluate the occupation number as 
\bea 
  \label{eq:gaugeoccno}
  f_{k}(t)
  ={1\over V} 
  \sqrt{ \langle |E_i(\vec k,t)|^2\rangle_\mathrm{cl}\langle |A_i^\mathrm{C}(\vec k,t)|^2 \rangle_\mathrm{cl}  }\,, 
\eea 
where we also use angle averaging within the $\langle \cdot\rangle_\textrm{cl}$ brackets to improve statistics, $\vec A^\mathrm{C}$ is the gauge field in Coulomb gauge, $\vec\nabla\cdot\vec A=0$, and $V$ is the volume.
The above expression can be conveniently calculated 
from the magnetic field using $ |\vec A^\mathrm{C}(\vec k,t)|^2= |\vec B(\vec k,t)|^2/k^2$ and thus does not require knowledge of the dispersion.
We have typically chosen a two-dimensional grid 
of $512^{2}$ to  $2048^{2}$ spatial points. 
Correspondingly, we consider the dispersion 
\bea
\label{eq:gaugedispersion}
   \omega^2_k 
   = { \langle |E_i(k) |^2 \rangle_\mathrm{cl} \over\langle |A_i^\mathrm{C}(k)|^2 \rangle_\mathrm{cl}} 
   = { \langle |E_i(k)|^2 \rangle_\mathrm{cl} k^2 \over\langle |B(k)|^2 \rangle_\mathrm{cl}}.
\eea  
We will also be interested in the two-point function of the Higgs field,
\begin{equation} \label{eq:cov}
 G(\vec x,\vec y,t)= \langle \phi(\vec x,t)  \phi(\vec y,t)^*\rangle_\mathrm{cl}\,,
\end{equation} 
but since this is not gauge invariant, it is not suitable for defining a meaningful occupation number. 
In contrast, 
\begin{equation} \label{eq:gaugecov}
 G^{U}(\vec x,\vec y,t) = \langle \phi(\vec x,t)U(\vec x,\vec y,t)  \phi(\vec y,t)^*\rangle_\mathrm{cl}\,,
\end{equation} 
(no angle averaging yet)
which involves the parallel transport operator
\begin{equation}\label{eq:U}
U(\vec x,\vec y,t)=\exp \left( i\, e\int_{\gamma({\vec x,\vec y})} d  \vec x' \cdot \vec A(\vec x',t)\right) \,
\end{equation}
along some path $\gamma({\vec x,\vec y})$ between $\vec x$ and $\vec y$, 
is a gauge invariant two-point function and thus better suited, despite the fact that it will have some residual path 
dependence. In practice, we use the zig-zag path 
along the lattice close to the straight line connecting 
$\vec x$  and $ \vec y $.

In analogy to the gauge sector we furthermore define the two-point correlation of the (gauge covariant) time derivative 
of the fields, 
\begin{equation} \label{eq:gaugecovDt}
 H^{U}(\vec x,\vec y,t) = \langle D_t \phi(\vec x,t)U(\vec x,\vec y,t) D_t \phi(\vec y,t)^*\rangle_\mathrm{cl}\,.
\end{equation} 
This serves to define the angle-averaged scalar field occupation number
($G^{U}(k,t)=\int d\Omega_{\mathbf{k}}G^{U}(\mathbf{k,t})$, etc.),
\bea \label{eq:scalarpnum}
  n_{k}(t)
  =  \sqrt{ G^U(k,t) H^U(k,t) },
\eea
and the dispersion,
\bea
\label{eq:Higgsdispersion}
\omega_k(t)= { H^U(k,t) \over G^U(k,t) },
\eea
in terms of the angle-averaged two-point functions $ H^U(k,t)$ and $G^U(k,t)$.

\subsection{Initial conditions}
We consider two different type of initial conditions and ranges of parameters.
The first is the `tachyonic' scenario, where we have $-\lambda v^{2}/6\equiv m^2<0$, and the expectation value of the Higgs field 
vanishes initially. 
For a purely scalar model, the evolution of such a system is well known to give rise to a non-zero expectation 
value of the Higgs field following  the tachyonic instability in which 
modes with $k < |m| $ become strongly 
populated \cite{Felder:2000hj,Asaka:2001ez,Sexty:2005pz}.

All modes are being populated with random amplitude and phase fluctuations to account for the quantum ground-state fluctuations 
with energy $\hbar\omega/2$.  
After the rescaling $ \Phi'=\sqrt{ \lambda} \Phi $ and using a
 small coupling this in practice gives field fluctuations 
with very small amplitude, and since the dynamics will be governed 
by the tachyonic instability, the actual value of the coupling 
(and thus the level of initial fluctuations) has very little impact on the 
dynamics. The negative mass-squared implies that the equilibrated system 
will be in the Higgs phase.

Second, in the `overpopulation' scenario, we choose the scalar modes to be occupied up to a cutoff $Q_s$ with a constant particle number
\bea 
n_k(t=0)= n_0 \theta ( Q_s - k ), 
\eea
with $ n_0 \lambda \simeq 1 $. We
choose $m^2=0$, such that the system will equilibrate to a state in
the Coulomb phase at late times, but as we will see below, the system 
shows a transient behavior exhibiting signs of being in the Higgs phase
before thermalisation. 
The initial gauge field modes are chosen empty. 
Hence, we expect early-time dynamics in the locally gauge invariant model that resembles that of a purely complex scalar field theory.

\begin{figure}[t]
\begin{center}
\epsfig{file=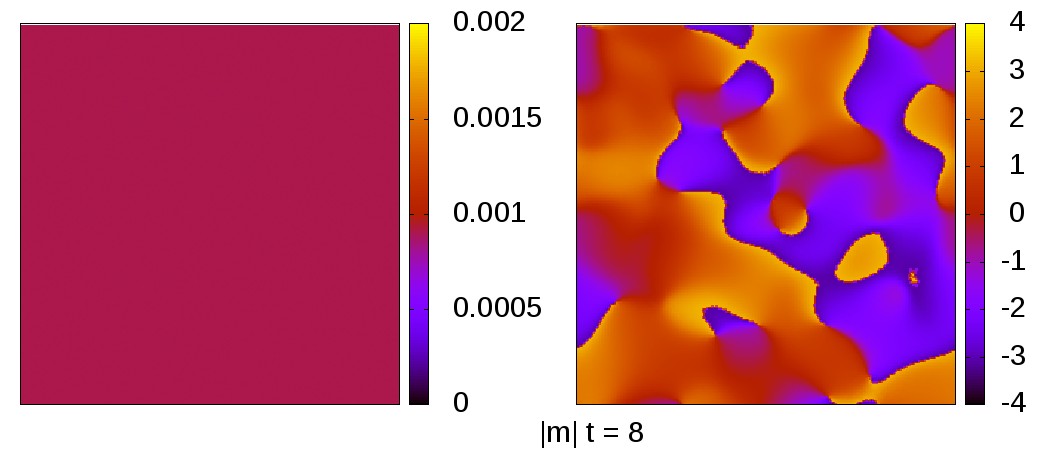, width=.95\columnwidth}
\epsfig{file=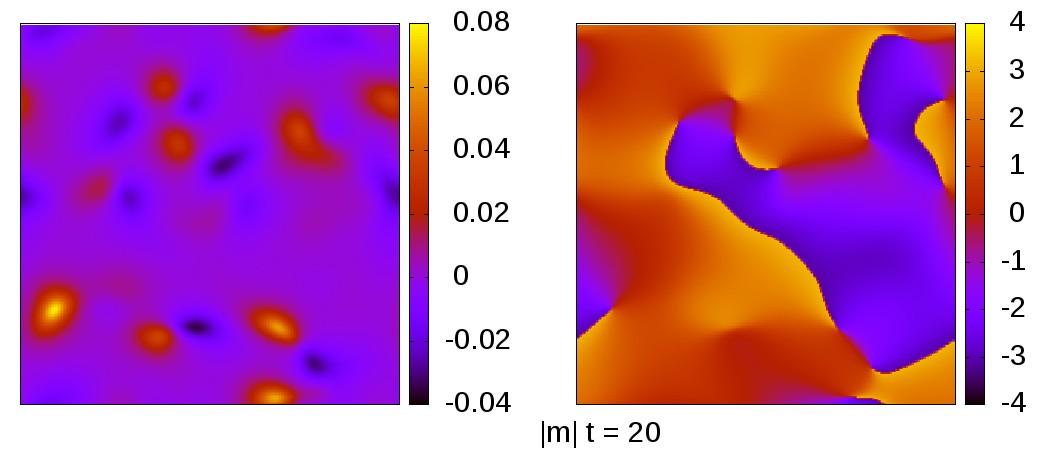, width=.95\columnwidth}
\epsfig{file=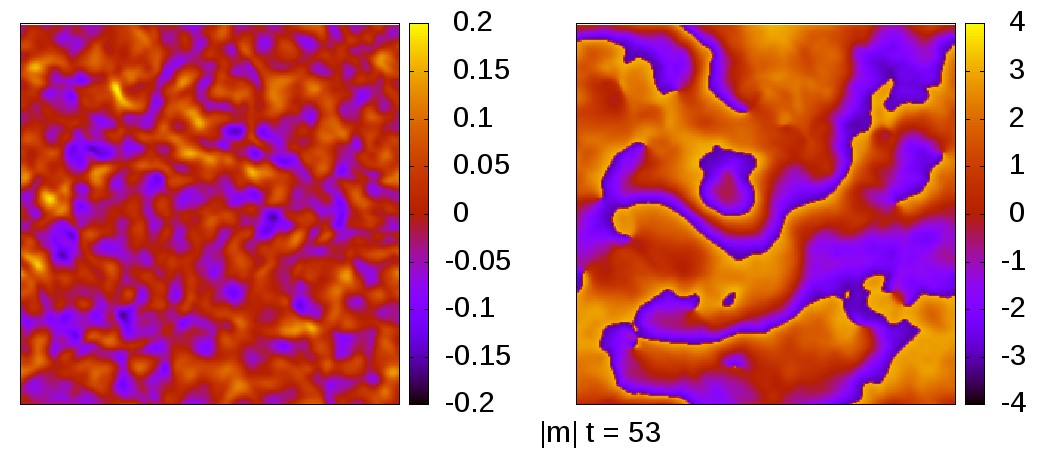, width=.95\columnwidth}
\caption{(Colour online) 
  Time evolution of the 2+1D system
  starting from `tachyonic' initial conditions, meaning that initially
  the expectation value of the Higgs field vanishes, with quantum
  noise in all momentum modes.  Each panel shows the entire $256^2$
  grid ($m^2a^2=0.05$) on a linear scale.  
  The otherwise periodic boundary conditions impose a net magnetic flux of $\Phi_{B}=2\pi/e$ through the area.
 The left column of panels shows the spatial
  magnetic field pattern, specifically, the dimensionless ratio
  $B(\vec x,t)/|m|^{3/2}$, at different times during the evolution ($ |m|t= 8,\
  20,\ 53 $, in the top, middle and bottom rows, respectively).  
  The right column shows the 
spatial configuration of the phase of the
  Higgs field in radians, encoded in colour between $-\pi$ and $\pi$.
  Vortex defects forming in the Higgs field at early times ($|m|t=8$)
  induce the formation of localized magnetic flux at intermediate
  times (center left panel).  At later times the smooth magnetic flux
  pattern destabilizes, short-range fluctuations build up, which also
  deteriorate the phase coherence of the Higgs field.  }
\label{fig:hdqcd1}
\end{center}
\end{figure}

\subsection{Vortex defects}
Pure scalar theories have recently been studied with respect to vortex formation and the interpretation of the corresponding field dynamics in terms of non-thermal fixed points \cite{Nowak:2010tm,Nowak:2011sk,Gasenzer:2011by,Schole:2012kt}.
The example of a non-relativistic Gross-Pitaevskii model served to illustrate that the evolution of a diluting vortex ensemble with vanishing total winding number corresponds to a self-similar process within which the system approaches a non-thermal fixed point, experiences critical slowing down and eventually moves away again from the critical point towards final thermalisation \cite{Schole:2012kt,Nowak:2012gd}.
The vortices appearing in the superfluid thereby take the important role of slowing down the evolution and stabilizing the field configuration against equilibration.
While scattering of vortices leads to a redistribution of the inter-vortex distances and initiates mutual defect annihilation, the final departure from the fixed point becomes possible only through the much weaker exchange of energy through sound modes propagating on the quasi-coherent background.

Coupling the evolving scalar field to a gauge sector, we expect the gauge field to become relevant at some stage during and after the initial evolution which resembles the dynamics of the pure scalar theory.
We specifically expect the scalar gauge field interaction to inflict the typical counter winding which is present in static solutions. 

In a pure scalar relativistic theory, Derrick's theorem \cite{Derrick1964a} forbids the formation of stable vortex solutions. 
In the Higgs phase, the above dynamic equations possess, however, stationary Abrikosov/Nielsen-Olesen vortex solutions \cite{Abrikosov1957a,Nielsen:1973cs} with spatially asymptotic behavior
\begin{equation}\label{eq:as}
\phi_\mathrm{as}(\vec x)= ve^{i\varphi} \0{(x_1+i\, x_2)^{n}}{|\vec x|}\,,
\quad   \vec A_{\rm as}  = \0{n}{i\, e} \vec \nabla \ln \phi_{\rm as}\,,
\end{equation}  
for $|\vec x|\to \infty$, with a constant phase $\varphi$ and winding number $n$, $|n|\ge1$.
The asymptotic form \eq{as} exhibits an important property of a vortex solution in a gauge theory. 
Asymptotically, the vortex winding carried by the scalar field is countered by that in the gauge
field. This is how Derrick's theorem is circumvented and a finite and even minimum energy for the vortex solution is arranged for. 
In the center of the vortex core the ansatz (\ref{eq:as}) is not valid 
anymore. 
The Higgs field has a zero at this point, the vortex configuration therefore
possesses a finite energy, and thus the appearance of dynamical vortices 
in scalar theories in ($1+d$)-dimensional theories with $d>1$ is possible.

\subsection{Magnetic confinement in the tachyonic scenario}
In \Sect{vortices+turbulence} we have discussed the occurrence and stability of vortices in the stationary limit and pointed to their importance for the scalar field dynamics. 
Starting from tachyonic initial conditions we find the scalar theory and the Higgs model to exhibit closely related dynamical evolution. 
In both cases, with and without gauge coupling, the phase distribution of the scalar field exhibits vortex defects with winding number $n=\pm1$.
The phase pattern shown in the right column of \Fig{hdqcd1} for a case with gauge coupling exhibits vortex defects, i.e., singular points around which the colour encoded phase angle wraps from $-\pi$ to $\pi$. 
The formation of these vortex-type configurations follows very similar patterns in the pure scalar and the gauge theory shown here.  

However, after the formation of the vortices in the scalar field, the dynamics differs. 
In the Higgs theory we find a reaction to occur in the magnetic field, forming typical magnetic Aharonov-Bohm-type fluxes
which in the stationary limit signal the formation of stable Nielsen-Olesen vortices in the coupled gauge-scalar system, as seen in the second row in \Fig{hdqcd1}. 
The respective left-column panels show the magnetic flux which, in the second row, is seen to become confined within the vortex cores.

During the progressing evolution, we find the confined magnetic flux to dissipate, however, due to the formation of additional vortex-antivortex pairs in the vicinity of the initial vortices.
These pairs seem to appear due to counter flow being induced by the rising magnetic flux and because the hybrid scalar-gauge vortex has not yet adapted to a Nielsen-Olesen equilibrium shape. 
Note that for a pure scalar vortex, the field modulus asymptotically approaches the bulk as $|\phi_\mathrm{bulk}|-|\phi|\sim1/r^{2}$ where $r$ is the distance from the vortex core, while for a Nielsen-Olesen vortex, both, the Higgs and the gauge field approach their bulk moduli exponentially in $r$.

Eventually, the system enters a phase of wildly fluctuating magnetic fields and a considerably changing phase of the Higgs field, see bottom row in \Fig{hdqcd1}.
This means, the intermediate magnetic confinement disappears.
The reason for this behavior is that the initial energy is too high to allow for a stable configuration bearing Nielsen-Olesen vortices.

To check whether the dynamics evolves to the expected equilibrium states we have also studied the evolution of a single scalar vortex with large winding number $n=5$.
We have found that it breaks up into five $n=1$ vortices for $\xi\lesssim\xi_\mathrm{LG}$ while it stays confined for $\xi\gtrsim\xi_\mathrm{LG}$ indicating a transition from an Abrikosov (type II) to a Meissner phase (type I) at $\xi_\mathrm{LG}\simeq0.5$. 
In both situations, type I and type II, the equilibrium magnetic field is confined in analogy to Meissner extrusion from the superconductor.

During the ensuing dynamical evolution the phase of the scalar field develops strong variations across the sample.
Considering, however, gauge invariant correlations, we find that the coherence present in the early-time field configuration is preserved at later times.
Specifically, the r\^{o}le of the covariance  \eq{cov}
in the scalar theory is taken over by its gauge invariant counter part defined in \Eq{gaugecov}.
One may speculate that the two observables behave similarly in both theories which would imply the possibility of quasi-universal similarities between them. 
This will be discussed more in the following section.

\subsection{Defects and electric confinement}
A particularly interesting question which arises with regard to previous results for scalar models \cite{Gasenzer:2011by} is how the dynamics occurring after the initial evolution due to the instability is ``distributed'' between the scalar and the gauge sectors.
In the following we will show that, for weak gauge coupling, i.e., a Landau-Ginzburg parameter $\xi\lesssim1/2$, the universal dynamics known to appear in a self-interacting Klein-Gordon model with global $U(1)$ symmetry is recovered in the locally gauge invariant theory.
It is found, in particular, that the pattern of the modulus of the scalar field as well as the gauge invariant pattern of relative phases between two given points in the system resembles the structure found in the purely scalar theory.
This implies the spontaneous spatial separation of regimes with opposite electric, i.e., Higgs charge, separated by sharp boundaries, without the occurrence of (meta)stable confined magnetic fluxes. 
As we will discuss further below, the appearance of this pattern can be related to the approach of a non-thermal fixed point of the dynamical evolution towards equilibrium.

\begin{figure}[t]
\begin{center}
\epsfig{file=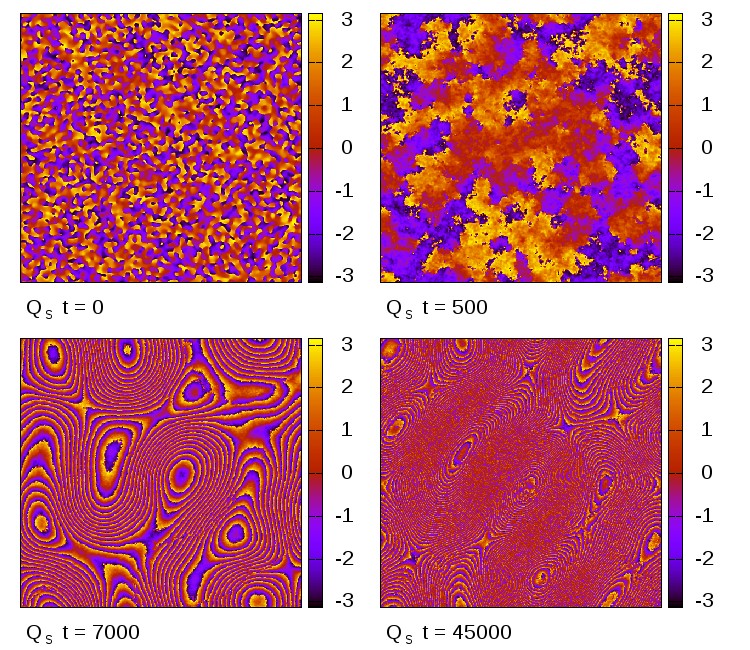, width=.95\columnwidth}
\caption{(Colour online) Time evolution of the spatial configuration of
  the phase angle $\varphi(\vec x,t)$ of the Higgs field $\phi=|\phi|\exp\{i\varphi\}$ on a $512^2$ lattice, starting from the
  overpopulation initial state in which all Higgs modes up to a
  maximum momentum scale $Q_{s}$ ($=0.5/a_s$ in lattice units) 
 are populated uniformly, each chosen with a random phase.  
 The simulation was done for $\xi =6e^{2}/\lambda= 0.025^2 $ in temporal gauge, $A_{0}=0$.  
 Times shown are $ Q_{s}t=0$, $500$,
  $7000$, $45000$, as indicated under each panel.  Each panel shows
  the colour-encoded phase angle in radians on the entire spatial grid
  on a linear scale.  The random initial phase becomes coherent within
  a short time, forming long-range coherent patterns which, at later
  times are superimposed by strong spatial phase gradients.  }
\label{fig:ScalarPhaseEvolution}
\end{center}
\end{figure}

\begin{figure}[t]
\begin{center}
\epsfig{file=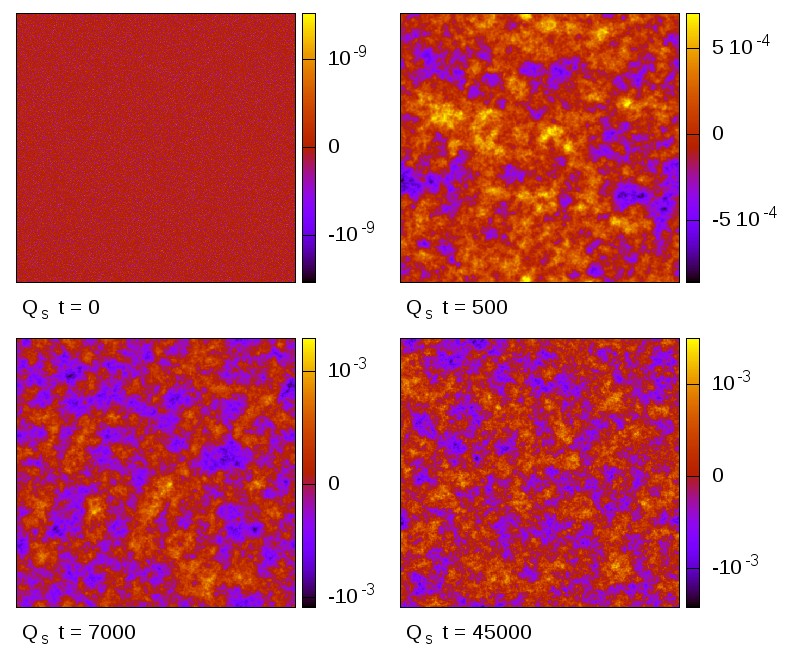, width=1.005\columnwidth}
\vspace{-2.6ex}
 \caption{(Colour online)
 The time evolution of the spatial configuration of the 
 dimensionless magnetic field strength $B(\vec x,t) / Q_s^{3/2} $, 
corresponding to the time evolution of the Higgs phase 
in \Fig{ScalarPhaseEvolution}. 
on a $512^2$ lattice using $ \xi = 0.025^2 $.
The panels show the same time steps as those in \Fig{ScalarPhaseEvolution}.
  }
\label{fig:BsquaredEvolution}
\end{center}
\end{figure}

\Fig{ScalarPhaseEvolution} shows the time evolution of the phase angle of the complex scalar field starting from the overpopulation initial condition, with $m^{2}=0$ and a weak gauge coupling $\xi=0.025^2 $ .
Similarly as in the tachyonic scenario discussed in the previous section, the pattern becomes soon dominated by strong phase rotations.
At the same time, the modulus squared of the magnetic field develops strong spatial fluctuations as is shown in a series of snapshots in \Fig{BsquaredEvolution}.

\begin{figure}[t]
\begin{center}
\epsfig{file=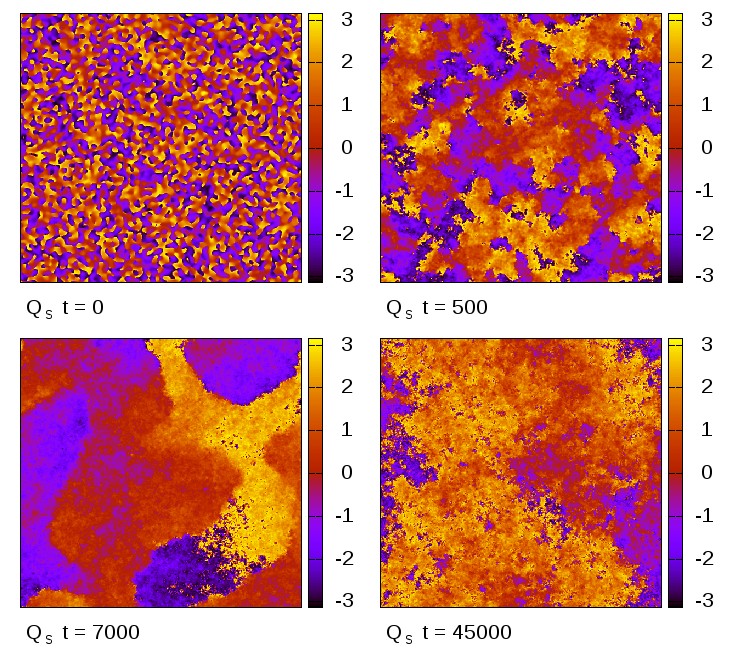, width=.94\columnwidth}
 \caption{(Colour online)
The spatial configuration of the relative phase $\varphi^{U}(\vec x,t)=\mathrm{arg}(G^U(\vec 0,\vec x,t))$ of 
the gauge covariant correlator $G^{U}$ where $\vec 0$ is taken to be 
the center of the lattice.
The colour-encoded phase is shown on the entire $512^2$ grid, on a linear scale, for $ \xi = 0.025^2 $. 
The panels show the same time steps as those in \Fig{ScalarPhaseEvolution}.
At short times, $Q_{s}t\lesssim 10^{3}$, the pattern is very close to that shown in \Fig{ScalarPhaseEvolution}, i.e., the build-up of long-range coherence in the Higgs field is not yet modified by the gauge potential. 
At the later stage, sharp boundaries appear in the gauge-covariant phase where the phase jumps by approximately $\pi$.
Along these defect lines the modulus squared of the Higgs field vanishes, cf.~\Fig{HiggsAbsvalueEvolution}.
The boundary lines separate regions of opposite phase rotation of $G^{U}$ in time and thus of opposite charge of the Higgs field.
 At late times the defects vanish and the system approaches a thermal configuration.
It is emphasized that the gauge-field contribution is relevant for the phase pattern in the lower two panels as is seen by comparing with \Fig{ScalarPhaseEvolution}.  
  }
\label{fig:Connphi}
\end{center}
\end{figure}

\begin{figure}[t]
\begin{center}
\epsfig{file=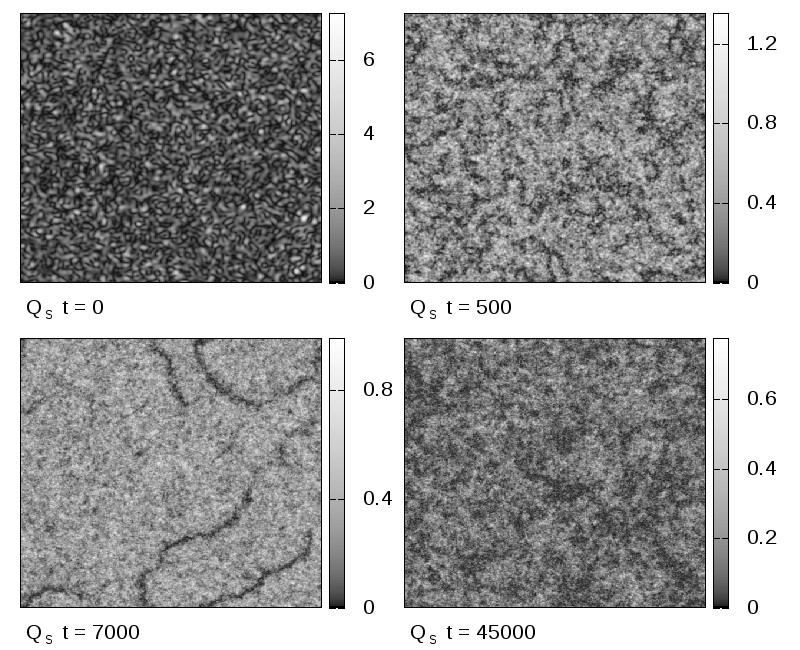, width=.95\columnwidth}
 \caption{
 The spatial configuration of the modulus squared of the Higgs field, specifically 
 the dimensionless combination $ \lambda |\phi|^4 / (6 Q_s^2 )$, on a $512^2$ lattice, for $ \xi = 0.025^2 $.
The panels show again the same time steps as those in \Fig{ScalarPhaseEvolution}.
The domain structure seen in the lower left panel corresponds to the domain structure in the gauge-invariant relative phase pattern seen at the corresponding time in \Fig{Connphi}.
At the black lines, the modulus of the Higgs field is close to zero due to the sharp phase kink of the gauge-invariant correlator $G^{U}$.
The lines separate regions of opposite phase rotation of $G^{U}$ in time and thus of opposite charge of the Higgs field, see \Fig{HiggsCharge}. 
  }
\label{fig:HiggsAbsvalueEvolution}
\end{center}
\end{figure}

\begin{figure}[t]
\begin{center}
\epsfig{file=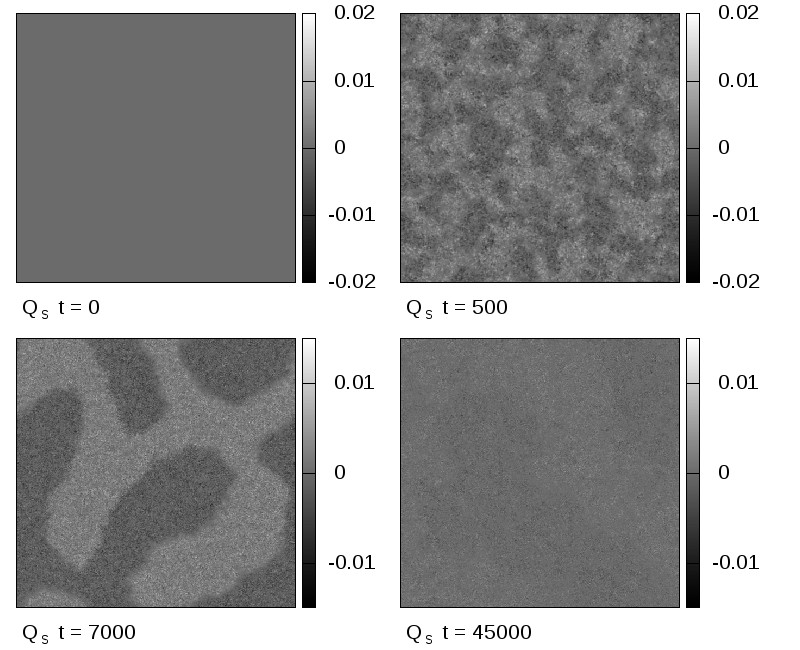, width=.99\columnwidth}
 \caption{
 The spatial configuration of the charge $J^{0}(\vec x, t)$ of the Higgs field, specifically 
 the dimensionless combination $J_0/(eQ_s^2)$, on a $512^2$ lattice, 
for $ \xi = 0.025^2 $.
The panels represent again the same time steps as those in \Fig{ScalarPhaseEvolution}.
This figure shows clearly the domain structure defined by the rotation sense of the phase of the Higgs field.
The phase kinks propagating along the boundaries are responsible for the long lifetime of the intermediate configuration near the non-thermal fixed point.
  }
\label{fig:HiggsCharge}
\end{center}
\end{figure}

On the contrary, depicting the evolution of the relative phase distribution as obtained from the gauge invariant correlator \eq{gaugecov} we find, at later times, much weaker variations, see \Fig{Connphi}.
The system rather quickly develops long-range phase coherence which is disrupted by defect lines separating large areas of almost equal phase.
The phase jumps by approximately $\pi$ at these defect lines.
Along these defect lines the modulus squared of the Higgs field is found to be suppressed to near zero, see \Fig{HiggsAbsvalueEvolution}.
The boundary lines separate regions of opposite phase rotation of $G^{U}$ in time and thus of \emph{opposite charge} of the Higgs field, see \Fig{HiggsCharge}, lower left panel, and the respective animations \cite{videos}.
In accordance with the observed charge distribution, inhomogeneous near-static electric fields build up.

Looking at the details of the time evolution we find that the phase jump by $\pi$ seen in \Fig{Connphi} is in fact the maximum of a phase jump oscillating sinusoidally in time.
This is the result of the oppositely rotating phases on either side of the boundary.
Due to an additional spatial oscillation of the phase along the lines, the observed phase jump and Higgs field suppression near this kink propagates in time along the boundary between the oppositely charged regions which are shown in \Fig{HiggsCharge}.
The propagating defect lines can be considered as elongated ``vortex sheets''.
These sheets have the form of phase defect lines of finite length, delimited by semi-vortex configurations at both ends.
The observed pattern is analogous to the formation of long-lived domains of opposite charge found in purely scalar simulations \cite{Gasenzer:2011by}.
 
It is emphasized that, here, the gauge-field pattern is releted to the phase pattern in the lower two panels as is seen by comparing with \Fig{ScalarPhaseEvolution}. 
The clear phase pattern is only seen in the gauge covariant quantities.
Differently stated, while the charge $J^{0}$, in our temporal gauge, is identical to the pure Higgs charge, the spatial current distribution $\vec J$ receives a distinct contribution from the gauge field $\vec A$.
As we will show in \Sect{StrongGaugeCoupl}, for stronger gauge coupling, $\xi\gtrsim1/2$, this gauge-field contribution leads to the deterioration of the observed long-range charge separation.

At late times, the charge difference gradually vanishes, see lower right panel of  \Fig{HiggsCharge}, and only much weaker fluctuations of the Higgs field around zero remain as is seen in  \Fig{HiggsAbsvalueEvolution}.
The system approaches a thermal configuration in the Coulomb phase as will be seen in the following results for momentum-space spectra.

Back to the intermediate-time pattern in the gauge field:
Given the almost flat, up to line defects, gauge covariant phase distribution of the Higgs field seen in the lower panels of \Fig{Connphi}, it becomes clear that the gradient of the phase distribution seen in \Fig{ScalarPhaseEvolution} reflects, at the later times (lower panels), the distribution of the vector potential $\vec A$ accross the sample.
Interestingly, this distribution shows clear vortex-type defects.
These, however, do not give rise to Aharonov-Bohm phases when integrated around a defect, and therefore do not correspond to type II Abrikosov  vortices enclosing magnetic flux.
Comparing with Figs.~\ref{fig:Connphi} and \ref{fig:HiggsCharge}, it becomes clear, however, that the position of these vortex defects is correlated with the pattern of the solitary defect lines which separate the regions of opposite charge, see also \cite{videos}.

In summary, at intermediate times of the equilibration process, a comparatively stable, slowly evolving defect structure is present in the gauge field.
It is clear that no smooth gauge transformation can be found to unwind the defects in the gauge field unless the vortices of opposite charge approach each other and mutually annihilate.
This annihilation process is not seen in our simulations. 
Comparing Figs.~\ref{fig:ScalarPhaseEvolution} and \ref{fig:HiggsCharge} we find that even at late times, when the electric confinement has disappeared, the vortices in the gauge potential remain, with equal numbers and density.
Comparing this to the cases of larger gauge coupling and thus $\xi$, just below as well as above the transition \cite{videos}, one finds that the density of vortices increases as does the size of the charged domains decrease.
For $\xi\gtrsim1/2$ there are no clearly separated defects visible any more.
These observations can be traced back to the slowly evolving electric fields between the regimes of opposite Higgs charge.
As we chose temporal gauge, $A_{0}=0$, the electric fields are proportional to the time derivative of the vector potential $\vec A$.
Hence, the spatial pattern of the vector potential reflects the time-integrated electric field pattern.
As the latter to first approximation only slowly evolves in time and since at late times it simply degrades in strength due to the charge difference between the regimes levelling off  \cite{Gasenzer:2011by} while keeping its pattern stable, see \Fig{HiggsCharge}, also the pattern in $\vec A$ survives to late times and conserves the previously existing structure.

We conclude that the vortex defects in the gauge field seen in our simulations do not directly give rise to physically visible magnetic confinement. 
They rather play an important part in the physics of electric confinement of nearly homogeneous charge distributions within sharply bounded regions.
In contrast to this, as discussed in the previous section, also Nielsen-Olesen vortices and anti-vortices, confining magnetic flux, are possible if the fluctuations in the short-wave-length modes are suppressed, i.e., there is considerably less energy in the system than in the realisations described here.
We will discuss, in the next section, that it is the formation of the charge-separated regimes which reflects the approach of a non-thermal fixed point, rather than the presence of a dilute ensemble of Nielsen-Olesen (anti-)vortices, as it is the case in a non-relativistic Gross-Pitaevskii system without gauge fields \cite{Nowak:2010tm,Nowak:2011sk}.

\begin{figure}
\begin{center}
\epsfig{file=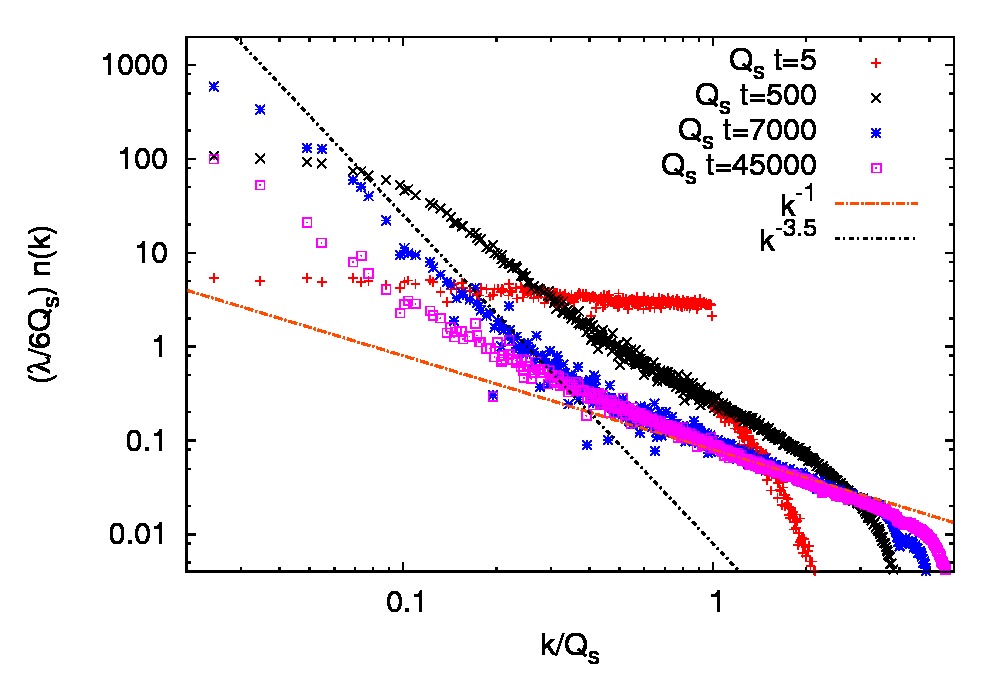, width=.95\columnwidth}
\vspace{-2ex}
 \caption{(Colour online) 
  Time evolution of the occupation number for the Higgs field defined in \Eq{scalarpnum}, multiplied by the dimensionless Higgs coupling $\lambda/6Q_{s}$, as a function of time, for  a coupling parameter $\xi=0.025^2 $.
 The different colours encode the time steps $Q_{s}t = 0, 1510, 11510, 39510$, with the $t=0$ configuration defining the overpopulation initial distribution.
 The approximate power law of $\sim k^{-3.5}$ indicates the approach of a non-thermal fixed point at intermediate times. 
 This power law coincides with the appearance of the domain structure as seen in position space in Figs.~\fig{Connphi} to \fig{HiggsCharge}. }
\label{fig:sksp_g0025}
\end{center}
\end{figure}

%
%
\subsection{Relation to a non-thermal fixed point}
The appearance of the above coherent spatial configurations is accompanied by characteristic gauge invariant power-law momentum distributions of the coupled gauge-scalar fields.
\Fig{sksp_g0025} shows the time evolution of the momentum distribution obtained by taking the Fourier transform of $G^{U}(\vec x,\vec y)$ with respect to $\vec x-\vec y$ and angle-averaging over the direction of these vectors.
Following the early-time overpopulation at intermediate momenta, together with an underpopulation at large $k$, the spectrum develops power-law regimes at intermediate times.
Comparing with the real-space patterns we find that this scaling, $n(k)\sim k^{-\zeta}$, with a characteristic infrared power $\zeta=3\dots3.5$ for $k\lesssim k_\mathrm{L}=0.3 Q_s$, corresponds to the onset of the formation of separate oppositely charged domains.
In particular, the extent of the scaling regime in the infrared is roughly of the order of the inverse size of the domains.
At large times, the domains gradually disappear, as does the infrared scaling, leaving an essentially thermalized classical ensemble of fluctuations which according to the Rayleigh-Jeans law scales as $n(k)\sim k^{-1}$.

In Refs.~\cite{Gasenzer:2011by,Karl:2013mn} the infrared scaling was shown to signal the approach of a non-thermal fixed point.
It was, in particular, argued in Ref.~\cite{Karl:2013mn} that the strongest infrared scaling $n(k)\sim k^{-\zeta}$ is related to the appearance of vortex-type defects.
The scaling thereby can be explained as arising from the geometric nature of the phase angle gradient around the vortex defect which falls off as $1/r$ as a function of the distance $r$ from the vortex core.
A non-integer power of $\zeta=3.5$ had been reported in Ref.~\cite{Gasenzer:2011by} which is likely to be explained as a mixed effect of an infrared power of $\zeta=4$ caused by vortices and $\zeta=3$ which indicates elongated soliton-like phase kinks as they are present in the domain walls.
In our case, the ensemble of ``vortex sheets'' is expected to have both effects, the steeper vortex-induced power law in the infrared due to the vortex-type behavior at the end of the sheet, in particular the contribution from short sheets and point-like defects, and the less steep power law due to the elongated, soliton-type nature of the sheets.

We remark that a dilute ensemble of Nielsen-Olesen vortices and anti-vortices, which is possible in the Higgs phase for low total energies, does not give rise to the infrared scaling seen for the non-gauged non-relativistic Gross-Pitaevskii model \cite{Nowak:2010tm,Nowak:2011sk} because the confined magnetic field and the Cooper current around the defects counteract each other.
This leads to a current which decays exponentially as a function of the distance $r$ from the core, as compared to the algebraic decay $\sim1/r$ around a Gross-Pitaevskii vortex.

We finally emphasize that the steep scaling with $\zeta=3$ (i.e., $\zeta=d+1$ in $d$ dimensions) was predicted within quantum field theory, extending kinetic-theory results from weak-wave-turbulence theory to the infrared regime where Boltzmann-type kinetic equations break down \cite{Berges:2008wm,Berges:2008sr,Scheppach:2009wu}.
Analyzing Kadanoff-Baym dynamic equations derived from a non-perturbatively resummed two-particle-irreducible (2PI) effective action the above power law was derived as the momentum-space signature of a non-thermal fixed point \cite{Berges:2008wm} and of strong wave turbulence in the infrared limit of strong occupation numbers. 

\subsection{Strong gauge coupling}
\label{sec:StrongGaugeCoupl}
\begin{figure}
\begin{center}
\epsfig{file=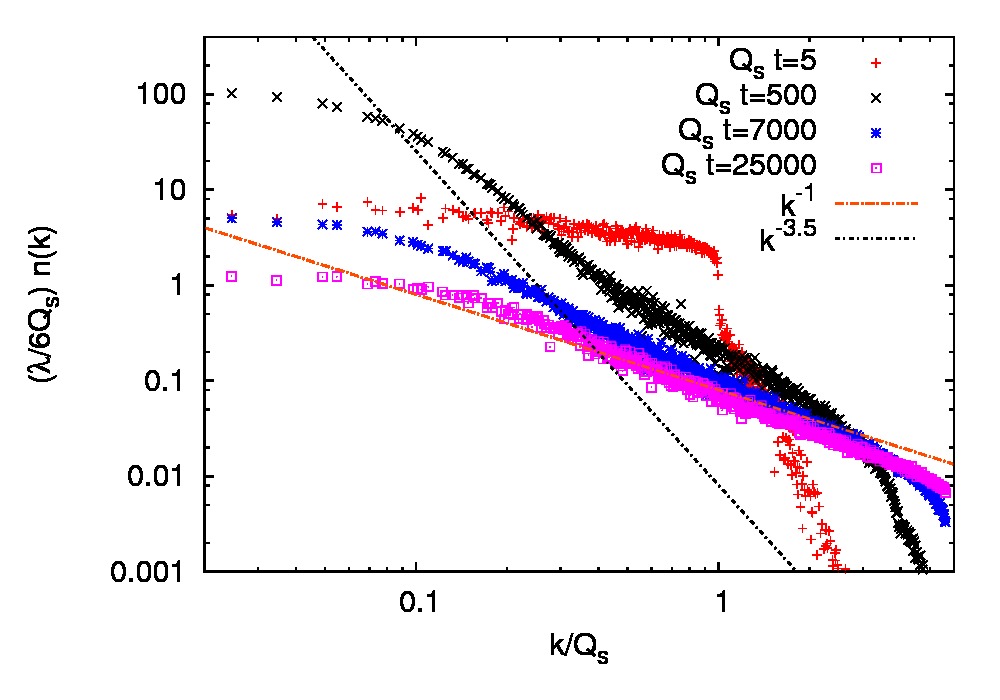, width=.95\columnwidth}
\epsfig{file=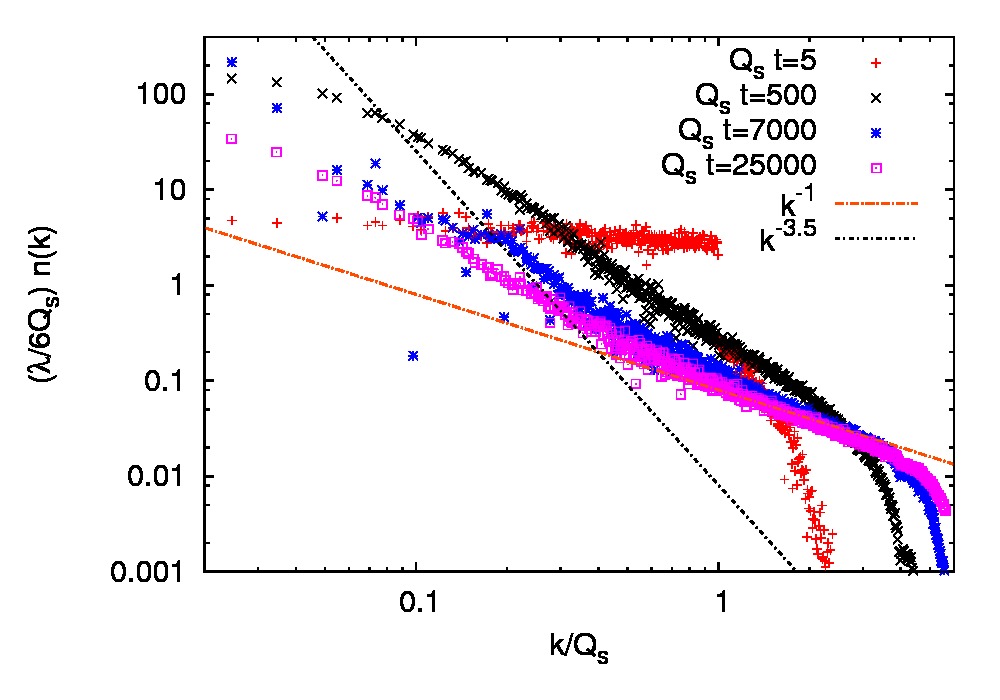, width=.95\columnwidth}
 \caption{(Colour online)  
 Same as in \Fig{sksp_g0025}, but for stronger gauge coupling $\xi=1$ (upper panel) 
and $\xi =0.25^2 $ (lower), at evolution times as indicated.
A scaling of $n(k)\sim k^{-3.5}$ as seen in \Fig{sksp_g0025} is less clear under the stronger gauge coupling chosen here.
}
\label{fig:GaugedMomDistrVarying-e}
\end{center}
\end{figure}

\Fig{GaugedMomDistrVarying-e} shows the evolution of the same momentum spectra as above for different gauge couplings $e$.
Increasing the gauge coupling, i.e., the Landau-Ginzburg parameter to $\xi>1/2$, which in equilibrium implies realisation of the type I Meissner phase, we find neither the characteristic infrared momentum scaling to appear during the chaotic evolution towards thermal equilibrium nor the clear domain formation in the gauge invariant spatial phase pattern.

\begin{figure}
\begin{center}
\epsfig{file=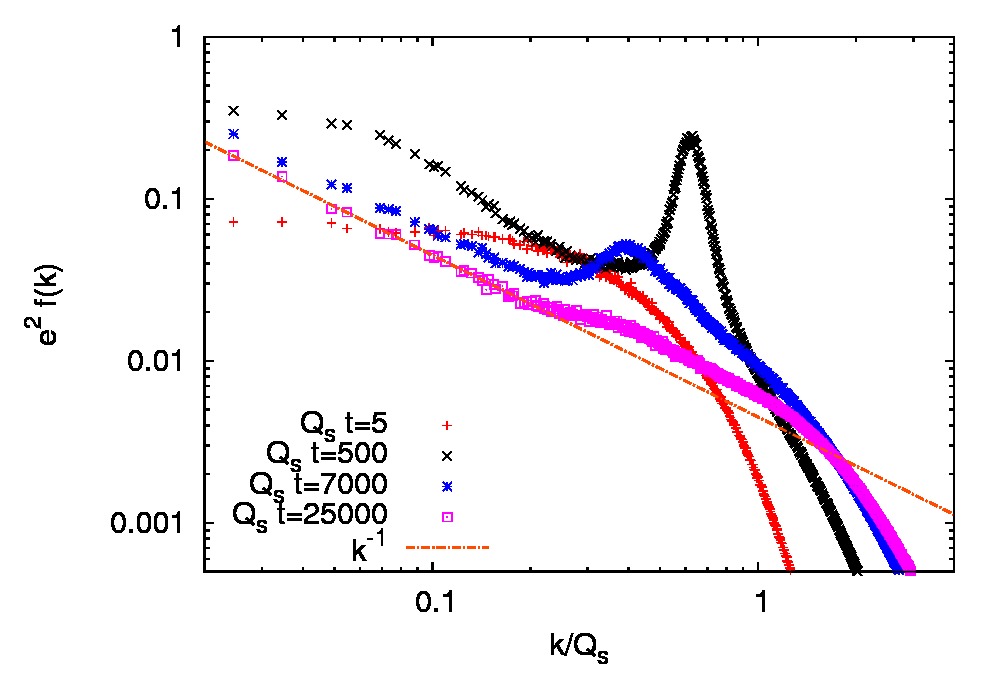, width=.95\columnwidth}
\epsfig{file=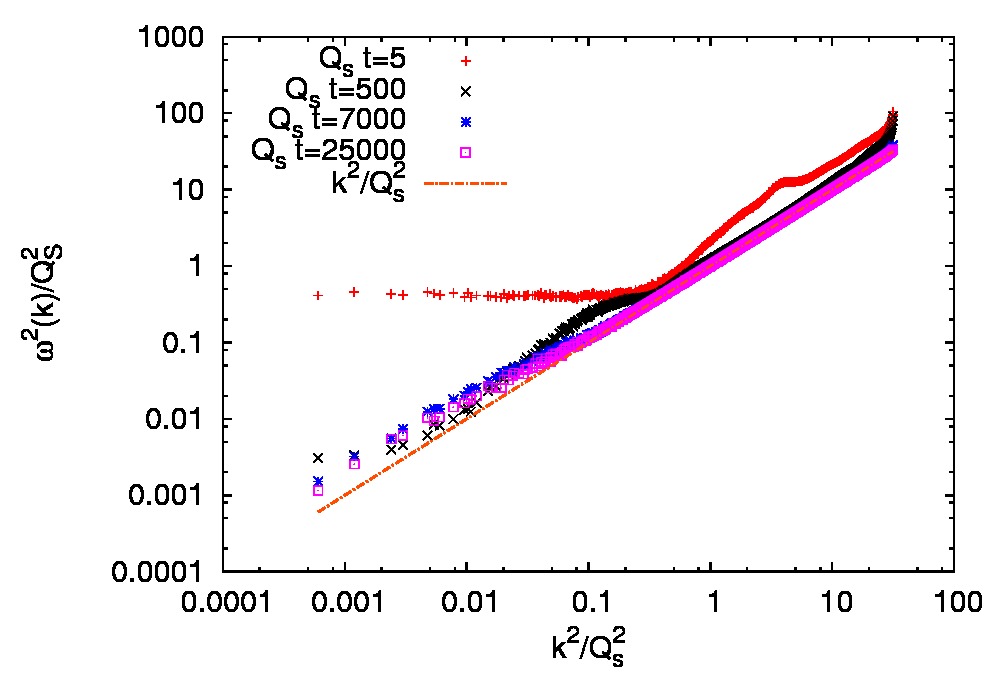, width=.95\columnwidth}
 \caption{(Colour online)  
 The time evolution of the occupation of the gauge-field modes as defined in \Eq{gaugeoccno} (upper panel), and their 
dispersion (\Eq{gaugedispersion}, lower panel)  starting from the overpopulation initial state, for $\xi=1$,  at evolution times as indicated.
}
\label{fig:gaugesp}
\end{center}
\end{figure}

The spectrum of the gauge fields for $\xi=1$ is shown in the upper panel of
Fig.~\ref{fig:gaugesp}. Since the gauge fields are unoccupied
initially, they are excited by the scalar fields, which have their dominant 
contribution to the energy density of the system at the scale $Q_s$.
One observes, at early times, a peak around $ k= 0.5\, Q_s$ in the gauge field 
spectrum, which slowly evolves into the thermal distribution of 
$n_k \sim k^{-1}$. 
In the lower panel, we also show the dispersion of the gauge modes, 
changing quickly from an initial massive behavior into 
the expected $ \omega_k \sim k$ for high modes. In the infrared, a discrepancy 
from the free dispersion is visible.

A gauge invariant and quasiparticle-definition \cite{Sexty:2005zm} 
independent way to 
characterize the phase of the system may be possible in terms 
of Wilson and t'Hooft loop variables, which are beyond the scope of this study,
or the screening properties of (static) gauge fields, as described in 
the next section.

\subsection{Screening dynamics of test charges}
As indicated in \Sect{Condensation} one can 
characterize the Higgs phase with strong electromagnetic fluctuations  by the screening of electric and magnetic charges. 
The screening behavior can be directly obtained by introducing 
test charges to the system. The test charges should be small enough such 
that they do not influence the physical state of the system.
We choose them, on the other hand, sufficiently large, such that the 
excess electric field can be detected clearly on the background
of the fluctuations in the plasma.

In the initial state, we introduce the test charges by solving 
the Poisson equation 
\bea \label{eq:poisson}
 \nabla\cdot \vec E_\mathrm{tc}(x,t=0) = \rho(\vec x) = c [\delta(\vec x-\vec x_1) -  \delta(\vec x-\vec x_2)],
\eea
adding a positive and a negative test charge  at positions $x_1$ and $x_2$, respectively,
such that the total charge in our box with periodical boundary conditions 
still vanishes. 
In the  $U(1)$-symmetric model, in which the gauge fields have no  
self interactions, the contribution of the initial plasma state
can be added to $E_\mathrm{tc}$ directly. 
For $SU(N)$ gauge fields, one 
has to solve the above equation for the full system, with 
the contribution of the plasma added to the r.h.s. of \Eq{poisson}.
The solution of \Eq{poisson} can be calculated easily by performing a
Fourier transformation on the lattice.

The solution of the EOM then supplies a time dependent
 electric field, 
the divergence of which satisfies
\begin{equation} \label{poisson2}
 \nabla\cdot \vec E_{tc}(\vec x,t=0) = c[ \delta(\vec x-\vec x_1) -  \delta(\vec x-\vec x_2)] + \rho(\vec x,t),
\end{equation}
where $\rho(x,t)$ is the charge density contribution 
of the Higgs field \eq{current}. This equation is satisfied 
automatically once the 
initial configuration is chosen to include a contribution according 
to \Eq{poisson}.  This is a consequence of the property of the EOM
that the external charge (defined below as the 'excess divergence' 
of the electric field) does not change as the fields evolve according to 
the EOM,  
\begin{eqnarray}
{ \partial \rho_{\rm ex}(\vec x,t) \over \partial t } &=&0\,,\\[2ex] 
\rho_{\rm ex}(\vec x,t) &=&\nabla\cdot \vec E(\vec x,t) - \rho_{\rm plasma}(\vec x,t).
\end{eqnarray}
Usually one uses the special case where $ \rho_{\rm ex}(\vec x,t)=0$, which 
just means there are no charges other than the particles in the 
plasma.
\begin{figure}[t]
\begin{center}
\epsfig{file=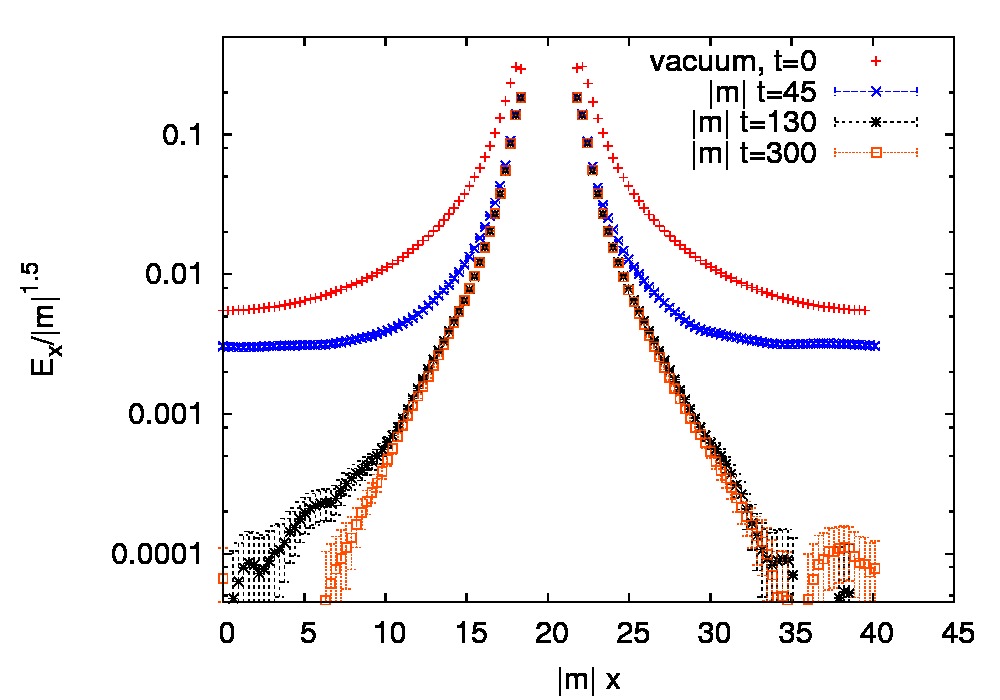, width=.95\columnwidth}
 \caption{(Colour online) 
  The spatial and temporal decay of the electric field around a testcharge, in the evolution starting from a tachyonic instability, with $ m^2a^2 = -0.1 $ on a $128^2$ lattice, for the fraction $ \xi=1$ of coupling constants.
}
\label{fig:electric_tachyonic}
\end{center}
\end{figure}

\begin{figure}
\begin{center}
\epsfig{file=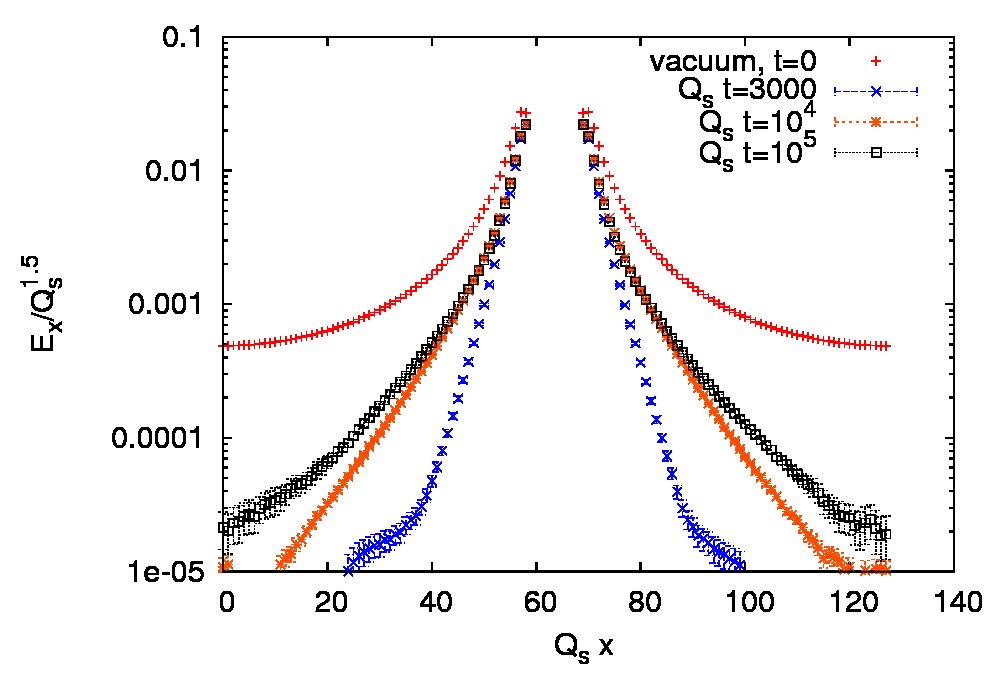, width=.95\columnwidth}
 \caption{(Colour online) 
  The spatial decay of the electric field around a testcharge, starting from the overpopulation 
initial configuration of the Higgs field and $m^2=0$.
The field is measured on a $128^2$ lattice with $ Q_sa=1$ and $\xi=1$. 
The field first decays in time before it raises again in the tails.}
\label{fig:electric_overpop}
\end{center}
\end{figure}

To measure the decay of the electric field, we insert two test charges 
of opposite sign onto the lattice. We then measure the 
component of the time-averaged electric field parallel to the line 
connecting the 
test charges, on the line connecting the two charges.
In \Fig{electric_tachyonic} we show the electric field of 
two test charges in the plasma after the Higgs field gets populated starting from the tachyonic initial condition. (Note that the time-averaged electric field 
points to the negative direction between 
the charges, therefore it does not show up on the plot.)
At $t=0$ we see the electric field 
of the charges in vacuum, described by the Coulomb law, while at later times 
we observe 
screening, i.e., the electric field decays exponentially with the distance. 
This is consistent with the expectation that the system 
ends up in the Higgs phase.

\Fig{electric_overpop} shows the electric field of the 
test charges developed from the overpopulation initial condition for the 
scalar fields, with initially unoccupied gauge fields, except for the field of 
the test charges.
We find that, despite the unscreened field at $t=0$, 
the electric fields become screened at intermediate times, while 
at late times, the field does not decay exponentially anymore implying that the 
fields are no longer screened.
This suggests a transient condensation 
of the scalar fields, as it has been seen for non-gauged 
systems \cite{Berges:2012us}, 
which leads to a screening of the electric fields at intermediate times. 
At late times when the condensate has decayed, the fields become again un-screened.
\begin{figure}[t]
\begin{center}
\epsfig{file=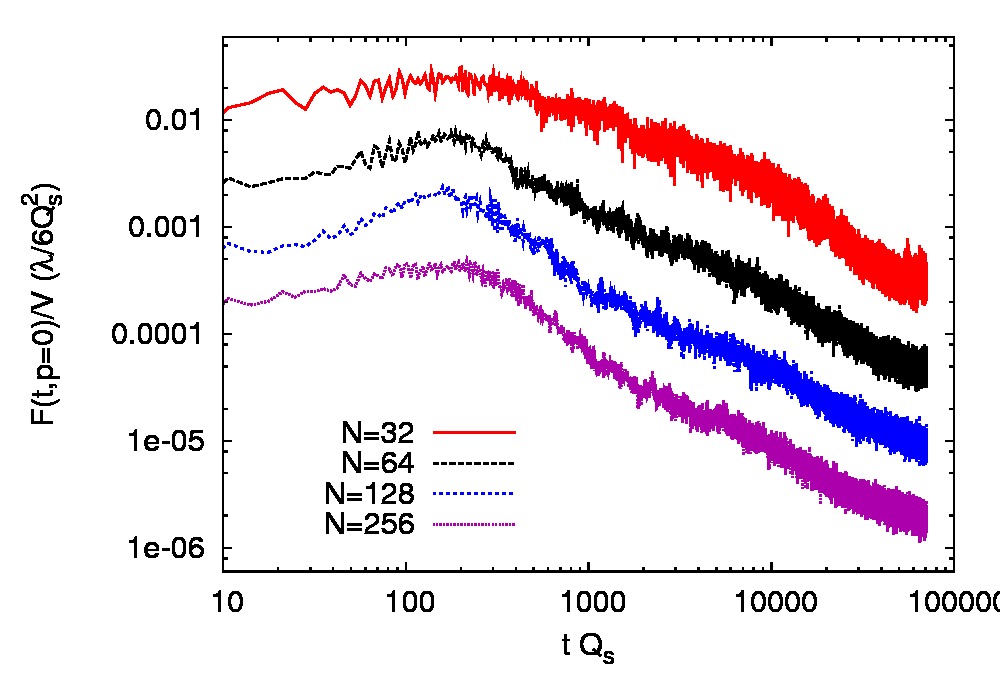, width=.95\columnwidth}
\epsfig{file=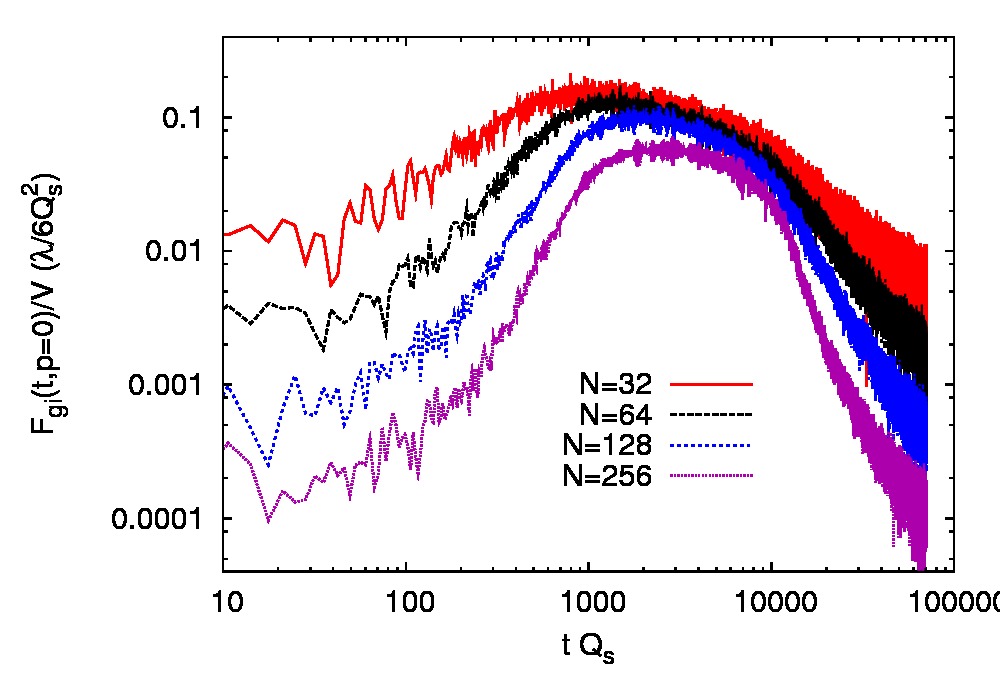, width=.95\columnwidth}
 \caption{(Colour online)  
 The time evolution of the zero-momentum mode of the gauge variant two point function $ F(p=0,t)$ (upper panel) and 
and its gauge invariant counter part $F_\mathrm{gi}(p=0,t)$ (lower panel), starting from 
an overpopulation initial condition, with $\xi=0.09 $, 
and $Q_s a= 0.707$, 
measured on $N^{2}$ lattices.
The lower panel demonstrates transient condensation of the system, as indicated by the overall volume independence of the zero-mode occupation within a time span from $Q_{s}t\simeq10^{3}$ to $10^{4}$.
}
\label{fig:cond03}
\end{center}
\end{figure}

\begin{figure}[t]
\begin{center}
\epsfig{file=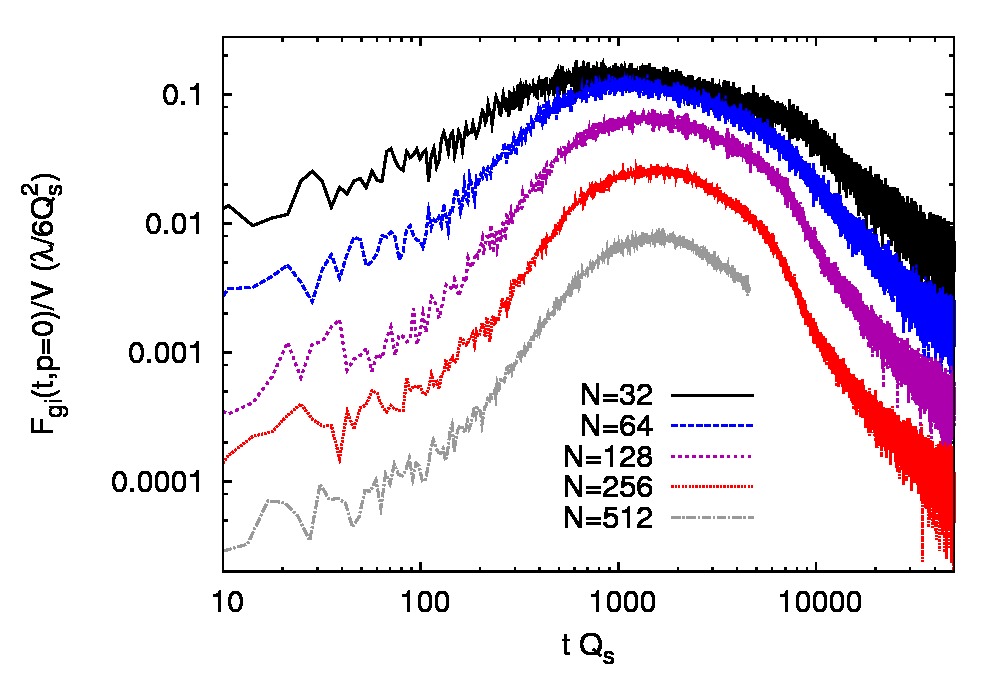,width=.95\columnwidth}
\epsfig{file=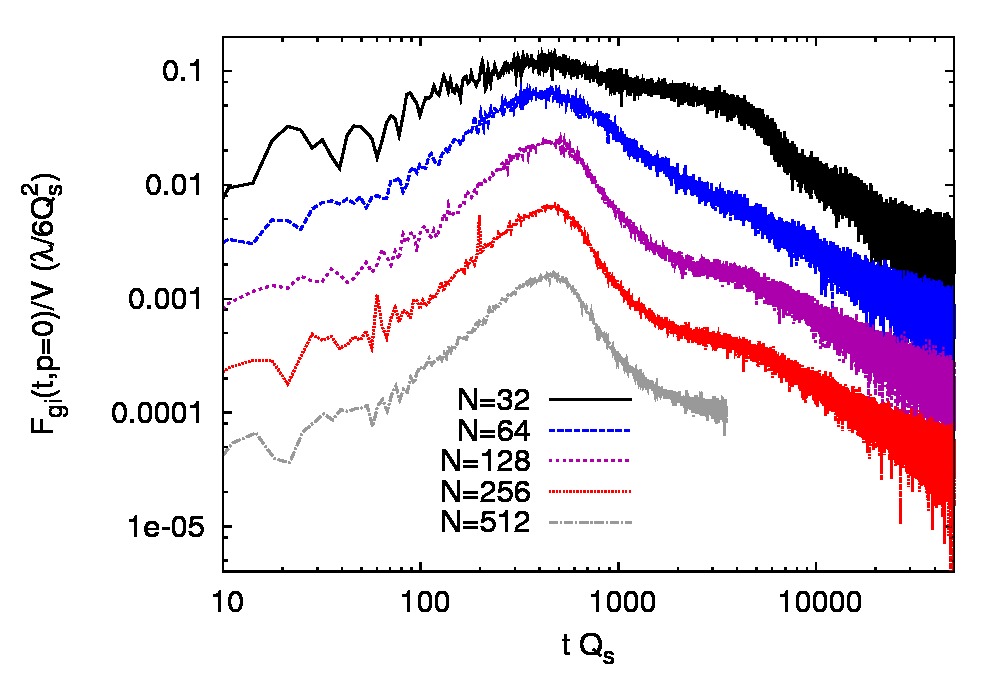,width=.95\columnwidth}
 \caption{ The gauge-invariant zero momentum two point 
functions $F_\mathrm{gi}(p=0,t)$ starting from 
an overpopulation initial condition using $ Q_s a= 0.707$, 
measured on $N^{2}$ lattices, at coupling $\xi=0.36 $ (upper panel), and  
$\xi=1$ (lower panel).
No transient condensation as is observed as in \Fig{cond03}.
}
\label{fig:cond079}
\end{center}
\end{figure}
\begin{figure}[t]
\begin{center}
\epsfig{file=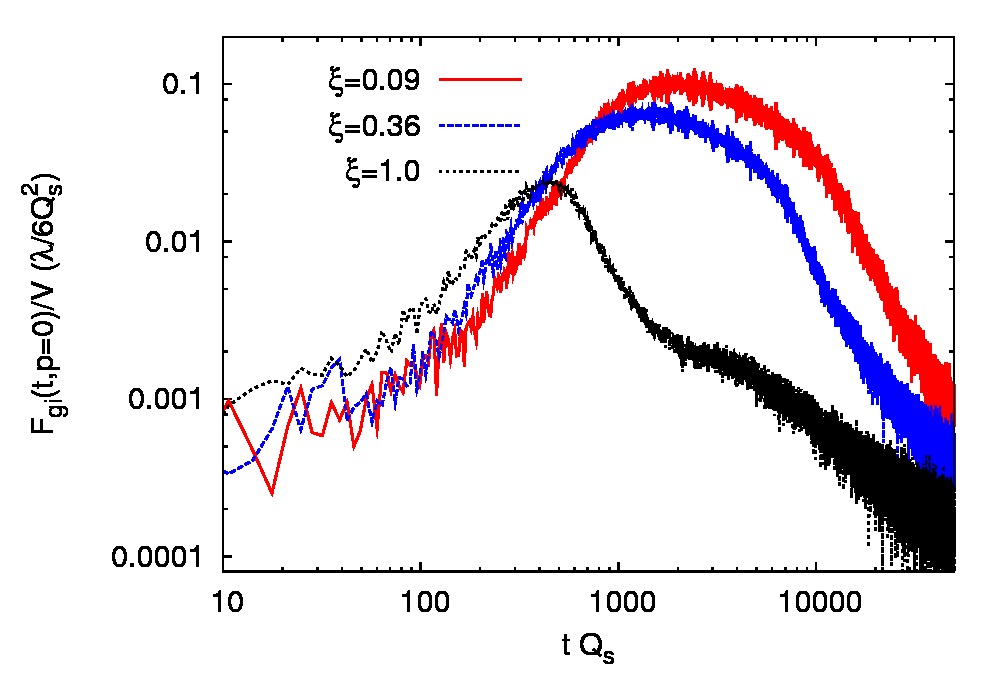,width=.95\columnwidth}
 \caption{ The gauge-invariant zero momentum two point 
functions $F_\mathrm{gi}(p=0,t)$ starting from 
an overpopulation initial condition using $ Q_s a= 0.7071$, 
measured on a $128^2$ lattice at different couplings, showing how condensation disappears for increased $\xi$.
}
\label{fig:cond0791}
\end{center}
\end{figure}

\subsection{Condensation}

The overpopulation initial state of a pure scalar field theory leads to 
non-equilibrium transient condensation even though the theory has a 
non-negative mass term \cite{Berges:2012us,Gasenzer:2011by}.
As we have seen in the previous section 
the screening properties of the electric fields suggest that 
the same phenomena also happen in a gauged system.
The question naturally arises whether one can see the analogous 
condensation in this system by studying the occupation 
numbers of the scalars directly. 

In the non-gauged scalar model,  condensation 
can be detected through the volume independence of the quantity
\begin{equation} \label{eq:ngi_cond}
 { F(p=0) \over V } = 
 { 1\over V^2 } \int d\vec x\, d\vec y\, G(\vec x,\vec y,t). 
\end{equation}

The local gauge symmetry allows the fields to have any direction in 
field space which means that, after averaging over gauge rotations, the 
above integral has only diagonal contributions left, i.e., those at $x=y$.
As a consequence, the integral is only proportional to the volume, and thus one
would argue that no condensation is possible.
We can make the above 
quantity gauge invariant by introducing a parallel transporter $U(x,y)$:
\bea \label{eq:gi_cond}
 { F^{U}(p=0) \over V } = 
 { 1\over V^2 } \int d\vec x\, d\vec y\, G^{U}(\vec x,\vec y,t) .
\eea
This corresponds to using the particle number 
definition in \Eq{scalarpnum}. This definition 
 introduces a path dependence which 
can become important when the gauge field fluctuations are strong.
In practice we choose 
the shortest possible path between the two end points, i.e., 
in general, a zig-zag path on the lattice.

With the gauge invariant definition of the 
two-point function, it is possible to recover 
the scenario which is realized for non-gauged scalars by 
considering  $\xi<1/2$.

In \Fig{cond03} the two-point functions introduced above are shown
for a simulation with $\xi=0.09 $. 
One observes that, while the observable \eq{ngi_cond} fails to 
show condensation, the gauge invariant quantity \eq{gi_cond} confirms
that transient condensation is present in the system. Note the relatively large deviation for the largest system size $N=256$.
This occurs because the coherence of the dynamics of the system is limited by the speed of light, and thus the build-up of coherence within a given time is only possible on scales smaller than the light propagation distance for that time. 
This means that exact volume independence is satisfied  only for small volumes. 
Larger volumes show a large growth, but the condensate starts to decay before the volume independence is reached.

In \Fig{cond079}, we show the gauge invariant two-point function
\eq{gi_cond} for couplings $\xi=0.36$ and $\xi=1.0 $. One observes 
that, as the photons become more relevant, the volume independence
of the condensate is no longer satisfied. It is remarkable that the quantity 
still shows a strong growth in the first part of the time evolution, but 
can not reach volume independence.  Whether this signals that condensation 
does not happen because of stronger decay processes, or whether the too strong 
contributions to \Eq{gi_cond} from gauge fields cause it to fail to signal condensation,
is an open question. 
Note that the zero-momentum two-point function shows a strong growth at short times 
suggesting a strong IR particle cascade, independent of the 
coupling $\xi$, see \Fig{cond0791}.

\section{Conclusions}

In this paper we have investigated the far-from-equilibrium dynamics of the Abelian Higgs model in two spatial 
dimensions. Using classical simulations we have studied initial conditions triggering a tachyonic instability 
as well as initial conditions with a strong overpopulation in the IR modes of the scalar field. 
We have compared the equilibration dynamics of the system for different choices of the Landau-Ginzburg parameter which, in equilibrium, determines whether the system in the phase of spontaneous $U(1)$-symmetry breaking is of type I (Meissner effect with magnetic fields expelled from the superconductor) or of type II (appearance of Abrikosov vortices within which confined magnetic flux can intrude the superconducting region).

For sufficiently weak gauge coupling, corresponding to a type II equilibrium Higgs phase, we find transient magnetic or electric confinement, topological defect formation in the gauge and Higgs fields, and turbulent scaling, even if the system at large times reaches an equilibrium state in the $U(1)$-symmetric or Coulomb phase.
We find, in particular, that electric charge separation appears in the system which has no net total charge, and that the regions with almost homogeneous charge density are separated by sharp soliton-like boundary walls.
These boundaries are caused by vortex sheets appearing in the gauge invariant Higgs phase correlator and are spatially correlated with vortex configurations in the gauge field which can not be removed by a smooth gauge transformation.
Hence, the topological structure present in the gauge field is crucially relevant for the appearance and dynamics of transient confinement in the electromagnetic quantities.

We have shown that the appearance of these defect structures is intimately related to the approach of a non-thermal fixed point \cite{Berges:2008wm}, showing up in universal infrared momentum scaling of the gauge-invariant Higgs correlation function.
This relation is consistent with earlier studies of non-equilibrium non-relativistic \cite{Nowak:2010tm,Nowak:2011sk,Schole:2012kt,Karl:2013mn} as well as relativistic \cite{Gasenzer:2011by} scalar field theories, in particular the dynamics of the defect formation for the approach of a non-thermal fixed point \cite{Schole:2012kt}.
The build-up of topological configurations corresponds to a shift of occupation numbers towards the infrared and causes an infrared strong-wave-turbulence cascade \cite{Scheppach:2009wu,Nowak:2012gd,Berges:2012us}.
We emphasise that the relation between the appearance of topological defects and non-thermal fixed points was also found in $3+1$ dimensional theories \cite{Nowak:2010tm,Nowak:2011sk,Gasenzer:2011by}.

We furthermore find transient Bose-Einstein condensation, proposed for a Glasma in Ref.~\cite{Blaizot:2011xf}. 
In Ref.~\cite{Berges:2012us}, such condensation was shown, 
for the un-gauged version of the 
theory we study here, to result from turbulent cascades.
According to our results, for weak gauge coupling $\xi=6e^{2}/\lambda\lesssim0.5$, using a gauge invariant definition of particle numbers \eq{scalarpnum}, one recovers the known behavior of the scalar theory. 

We have also studied the screening of static electric fields as a signal 
of the transient behavior in the system by inserting test-charges into
the plasma and measuring the decay of their fields. The results 
suggest that the overpopulation scenario indeed leads to
a transient condensation and a non-thermal fixed point being approached.

In summary, condensation, turbulence, the appearance of topological defects in the gauge fields, and confinement are found to be all closely related.

\section{Acknowledgements}
We thank J. Berges, M. Karl and S. Schlichting for discussions. The research of L. McLerran is
supported under DOE Contract No. DE-AC02-98CH10886.  Larry McLerran
gratefully acknowledges support as a Hans Jensen Professor at the
Institute for Theoretical Physics, University of Heidelberg, and the
hospitality of this Institute where this work was initiated.  This
work is supported by Deutsche Forschungsgemeinschaft (GA677/7,8), the
Helmholtz Alliance HA216/EMMI and by ERC-AdG-290623. Parts of the
numerical calculations were done on the bwGRiD
(http://www.bw-grid.de), member of the German D-Grid initiative,
funded by BMBF and MWFK Baden-W\"urttemberg.

\noindent 


\begin{thebibliography}{55}
\expandafter\ifx\csname natexlab\endcsname\relax\def\natexlab#1{#1}\fi
\expandafter\ifx\csname bibnamefont\endcsname\relax
  \def\bibnamefont#1{#1}\fi
\expandafter\ifx\csname bibfnamefont\endcsname\relax
  \def\bibfnamefont#1{#1}\fi
\expandafter\ifx\csname citenamefont\endcsname\relax
  \def\citenamefont#1{#1}\fi
\expandafter\ifx\csname url\endcsname\relax
  \def\url#1{\texttt{#1}}\fi
\expandafter\ifx\csname urlprefix\endcsname\relax\def\urlprefix{URL }\fi
\providecommand{\bibinfo}[2]{#2}
\providecommand{\eprint}[2][]{\url{#2}}

\bibitem[{\citenamefont{Kovner et~al.}(1995)\citenamefont{Kovner, McLerran, and
  Weigert}}]{Kovner:1995ts}
\bibinfo{author}{\bibfnamefont{A.}~\bibnamefont{Kovner}},
  \bibinfo{author}{\bibfnamefont{L.~D.} \bibnamefont{McLerran}},
  \bibnamefont{and} \bibinfo{author}{\bibfnamefont{H.}~\bibnamefont{Weigert}},
  \bibinfo{journal}{Phys. Rev. D} \textbf{\bibinfo{volume}{52}},
  \bibinfo{pages}{3809} (\bibinfo{year}{1995}), \eprint{hep-ph/9505320}.

\bibitem[{\citenamefont{Krasnitz and Venugopalan}(1999)}]{Krasnitz:1998ns}
\bibinfo{author}{\bibfnamefont{A.}~\bibnamefont{Krasnitz}} \bibnamefont{and}
  \bibinfo{author}{\bibfnamefont{R.}~\bibnamefont{Venugopalan}},
  \bibinfo{journal}{Nucl. Phys.} \textbf{\bibinfo{volume}{B557}},
  \bibinfo{pages}{237} (\bibinfo{year}{1999}), \eprint{hep-ph/9809433}.

\bibitem[{\citenamefont{Krasnitz et~al.}(2003)\citenamefont{Krasnitz, Nara, and
  Venugopalan}}]{Krasnitz:2002mn}
\bibinfo{author}{\bibfnamefont{A.}~\bibnamefont{Krasnitz}},
  \bibinfo{author}{\bibfnamefont{Y.}~\bibnamefont{Nara}}, \bibnamefont{and}
  \bibinfo{author}{\bibfnamefont{R.}~\bibnamefont{Venugopalan}},
  \bibinfo{journal}{Nucl. Phys.} \textbf{\bibinfo{volume}{A717}},
  \bibinfo{pages}{268} (\bibinfo{year}{2003}), \eprint{hep-ph/0209269}.

\bibitem[{\citenamefont{Lappi}(2003)}]{Lappi:2003bi}
\bibinfo{author}{\bibfnamefont{T.}~\bibnamefont{Lappi}},
  \bibinfo{journal}{Phys. Rev. C} \textbf{\bibinfo{volume}{67}},
  \bibinfo{pages}{054903} (\bibinfo{year}{2003}), \eprint{hep-ph/0303076}.

\bibitem[{\citenamefont{Lappi and McLerran}(2006)}]{Lappi:2006fp}
\bibinfo{author}{\bibfnamefont{T.}~\bibnamefont{Lappi}} \bibnamefont{and}
  \bibinfo{author}{\bibfnamefont{L.}~\bibnamefont{McLerran}},
  \bibinfo{journal}{Nucl. Phys.} \textbf{\bibinfo{volume}{A772}},
  \bibinfo{pages}{200} (\bibinfo{year}{2006}), \eprint{hep-ph/0602189}.

\bibitem[{\citenamefont{Gribov et~al.}(1983)\citenamefont{Gribov, Levin, and
  Ryskin}}]{Gribov:1984tu}
\bibinfo{author}{\bibfnamefont{L.}~\bibnamefont{Gribov}},
  \bibinfo{author}{\bibfnamefont{E.}~\bibnamefont{Levin}}, \bibnamefont{and}
  \bibinfo{author}{\bibfnamefont{M.}~\bibnamefont{Ryskin}},
  \bibinfo{journal}{Phys. Rept.} \textbf{\bibinfo{volume}{100}},
  \bibinfo{pages}{1} (\bibinfo{year}{1983}).

\bibitem[{\citenamefont{Mueller and Qiu}(1986)}]{Mueller:1985wy}
\bibinfo{author}{\bibfnamefont{A.~H.} \bibnamefont{Mueller}} \bibnamefont{and}
  \bibinfo{author}{\bibfnamefont{J.-w.} \bibnamefont{Qiu}},
  \bibinfo{journal}{Nucl. Phys.} \textbf{\bibinfo{volume}{B268}},
  \bibinfo{pages}{427} (\bibinfo{year}{1986}).

\bibitem[{\citenamefont{McLerran and
  Venugopalan}(1994{\natexlab{a}})}]{McLerran:1993ni}
\bibinfo{author}{\bibfnamefont{L.~D.} \bibnamefont{McLerran}} \bibnamefont{and}
  \bibinfo{author}{\bibfnamefont{R.}~\bibnamefont{Venugopalan}},
  \bibinfo{journal}{Phys. Rev. D} \textbf{\bibinfo{volume}{49}},
  \bibinfo{pages}{2233} (\bibinfo{year}{1994}{\natexlab{a}}),
  \eprint{hep-ph/9309289}.

\bibitem[{\citenamefont{McLerran and
  Venugopalan}(1994{\natexlab{b}})}]{McLerran:1993ka}
\bibinfo{author}{\bibfnamefont{L.~D.} \bibnamefont{McLerran}} \bibnamefont{and}
  \bibinfo{author}{\bibfnamefont{R.}~\bibnamefont{Venugopalan}},
  \bibinfo{journal}{Phys. Rev. D} \textbf{\bibinfo{volume}{49}},
  \bibinfo{pages}{3352} (\bibinfo{year}{1994}{\natexlab{b}}),
  \eprint{hep-ph/9311205}.

\bibitem[{\citenamefont{Blaizot et~al.}(2012)\citenamefont{Blaizot, Gelis,
  Liao, McLerran, and Venugopalan}}]{Blaizot:2011xf}
\bibinfo{author}{\bibfnamefont{J.-P.} \bibnamefont{Blaizot}},
  \bibinfo{author}{\bibfnamefont{F.}~\bibnamefont{Gelis}},
  \bibinfo{author}{\bibfnamefont{J.-F.} \bibnamefont{Liao}},
  \bibinfo{author}{\bibfnamefont{L.}~\bibnamefont{McLerran}}, \bibnamefont{and}
  \bibinfo{author}{\bibfnamefont{R.}~\bibnamefont{Venugopalan}},
  \bibinfo{journal}{Nucl. Phys.} \textbf{\bibinfo{volume}{A873}},
  \bibinfo{pages}{68} (\bibinfo{year}{2012}), \eprint{1107.5296}.

\bibitem[{\citenamefont{Kurkela and Moore}(2011)}]{Kurkela:2011ti}
\bibinfo{author}{\bibfnamefont{A.}~\bibnamefont{Kurkela}} \bibnamefont{and}
  \bibinfo{author}{\bibfnamefont{G.~D.} \bibnamefont{Moore}},
  \bibinfo{journal}{JHEP} \textbf{\bibinfo{volume}{1112}}, \bibinfo{pages}{044}
  (\bibinfo{year}{2011}), \eprint{1107.5050}.

\bibitem[{\citenamefont{Berges et~al.}(2012)\citenamefont{Berges, Schlichting,
  and Sexty}}]{Berges:2012ev}
\bibinfo{author}{\bibfnamefont{J.}~\bibnamefont{Berges}},
  \bibinfo{author}{\bibfnamefont{S.}~\bibnamefont{Schlichting}},
  \bibnamefont{and} \bibinfo{author}{\bibfnamefont{D.}~\bibnamefont{Sexty}},
  \bibinfo{journal}{Phys. Rev. D} \textbf{\bibinfo{volume}{86}},
  \bibinfo{pages}{074006} (\bibinfo{year}{2012}), \eprint{1203.4646}.

\bibitem[{\citenamefont{Berges et~al.}(2013)\citenamefont{Berges, Boguslavski,
  Schlichting, and Venugopalan}}]{Berges:2013eia}
\bibinfo{author}{\bibfnamefont{J.}~\bibnamefont{Berges}},
  \bibinfo{author}{\bibfnamefont{K.}~\bibnamefont{Boguslavski}},
  \bibinfo{author}{\bibfnamefont{S.}~\bibnamefont{Schlichting}},
  \bibnamefont{and}
  \bibinfo{author}{\bibfnamefont{R.}~\bibnamefont{Venugopalan}}
  (\bibinfo{year}{2013}), \eprint{1303.5650}.

\bibitem[{\citenamefont{Svistunov}(1991)}]{Svistunov1991a}
\bibinfo{author}{\bibfnamefont{B.}~\bibnamefont{Svistunov}},
  \bibinfo{journal}{J. Mosc. Phys. Soc.} \textbf{\bibinfo{volume}{1}},
  \bibinfo{pages}{373} (\bibinfo{year}{1991}).

\bibitem[{\citenamefont{Berloff and Svistunov}(2002)}]{Berloff2002a}
\bibinfo{author}{\bibfnamefont{N.~G.} \bibnamefont{Berloff}} \bibnamefont{and}
  \bibinfo{author}{\bibfnamefont{B.~V.} \bibnamefont{Svistunov}},
  \bibinfo{journal}{Phys. Rev. A} \textbf{\bibinfo{volume}{66}},
  \bibinfo{pages}{013603} (\bibinfo{year}{2002}).

\bibitem[{\citenamefont{Berges and Sexty}(2012)}]{Berges:2012us}
\bibinfo{author}{\bibfnamefont{J.}~\bibnamefont{Berges}} \bibnamefont{and}
  \bibinfo{author}{\bibfnamefont{D.}~\bibnamefont{Sexty}},
  \bibinfo{journal}{Phys. Rev. Lett.} \textbf{\bibinfo{volume}{108}},
  \bibinfo{pages}{161601} (\bibinfo{year}{2012}), \eprint{1201.0687}.

\bibitem[{\citenamefont{Nowak and Gasenzer}(2012)}]{Nowak:2012gd}
\bibinfo{author}{\bibfnamefont{B.}~\bibnamefont{Nowak}} \bibnamefont{and}
  \bibinfo{author}{\bibfnamefont{T.}~\bibnamefont{Gasenzer}},
  \bibinfo{journal}{arXiv: 1206.3181 [cond-mat.quant-gas]}
  (\bibinfo{year}{2012}), \eprint{1206.3181}.

\bibitem[{\citenamefont{Frisch}(1995)}]{Frisch1995a}
\bibinfo{author}{\bibfnamefont{U.}~\bibnamefont{Frisch}},
  \emph{\bibinfo{title}{Turbulence: The Legacy of A. N. Kolmogorov}}
  (\bibinfo{publisher}{CUP, Cambridge, UK}, \bibinfo{year}{1995}).

\bibitem[{\citenamefont{Zakharov et~al.}(1992)\citenamefont{Zakharov, {L'vov},
  and Falkovich}}]{Zakharov1992a}
\bibinfo{author}{\bibfnamefont{V.~E.} \bibnamefont{Zakharov}},
  \bibinfo{author}{\bibfnamefont{V.~S.} \bibnamefont{{L'vov}}},
  \bibnamefont{and}
  \bibinfo{author}{\bibfnamefont{G.}~\bibnamefont{Falkovich}},
  \emph{\bibinfo{title}{Kolmogorov Spectra of Turbulence I: Wave Turbulence}}
  (\bibinfo{publisher}{Springer, Berlin}, \bibinfo{year}{1992}).

\bibitem[{\citenamefont{Nazarenko}(2011)}]{Nazarenko2011a}
\bibinfo{author}{\bibfnamefont{S.}~\bibnamefont{Nazarenko}},
  \emph{\bibinfo{title}{Wave turbulence}}, no. \bibinfo{number}{825} in
  \bibinfo{series}{Lecture Notes in Physics} (\bibinfo{publisher}{Springer},
  \bibinfo{address}{Heidelberg}, \bibinfo{year}{2011}).

\bibitem[{\citenamefont{Berges et~al.}(2009)\citenamefont{Berges, Scheffler,
  and Sexty}}]{Berges:2008mr}
\bibinfo{author}{\bibfnamefont{J.}~\bibnamefont{Berges}},
  \bibinfo{author}{\bibfnamefont{S.}~\bibnamefont{Scheffler}},
  \bibnamefont{and} \bibinfo{author}{\bibfnamefont{D.}~\bibnamefont{Sexty}},
  \bibinfo{journal}{Phys. Lett.} \textbf{\bibinfo{volume}{B681}},
  \bibinfo{pages}{362} (\bibinfo{year}{2009}), \eprint{0811.4293}.

\bibitem[{\citenamefont{Fukushima and Gelis}(2012)}]{Fukushima:2011nq}
\bibinfo{author}{\bibfnamefont{K.}~\bibnamefont{Fukushima}} \bibnamefont{and}
  \bibinfo{author}{\bibfnamefont{F.}~\bibnamefont{Gelis}},
  \bibinfo{journal}{Nucl. Phys.} \textbf{\bibinfo{volume}{A874}},
  \bibinfo{pages}{108} (\bibinfo{year}{2012}), \eprint{1106.1396}.

\bibitem[{\citenamefont{Schlichting}(2012)}]{Schlichting:2012es}
\bibinfo{author}{\bibfnamefont{S.}~\bibnamefont{Schlichting}},
  \bibinfo{journal}{Phys. Rev. D} \textbf{\bibinfo{volume}{86}},
  \bibinfo{pages}{065008} (\bibinfo{year}{2012}), \eprint{1207.1450}.

\bibitem[{\citenamefont{Kurkela and Moore}(2012)}]{Kurkela:2012hp}
\bibinfo{author}{\bibfnamefont{A.}~\bibnamefont{Kurkela}} \bibnamefont{and}
  \bibinfo{author}{\bibfnamefont{G.~D.} \bibnamefont{Moore}},
  \bibinfo{journal}{Phys. Rev. D} \textbf{\bibinfo{volume}{86}},
  \bibinfo{pages}{056008} (\bibinfo{year}{2012}), \eprint{1207.1663}.

\bibitem[{\citenamefont{Fukushima}(2013)}]{Fukushima:2013dma}
\bibinfo{author}{\bibfnamefont{K.}~\bibnamefont{Fukushima}}
  (\bibinfo{year}{2013}), \eprint{1307.1046}.

\bibitem[{\citenamefont{Nowak et~al.}(2011)\citenamefont{Nowak, Sexty, and
  Gasenzer}}]{Nowak:2010tm}
\bibinfo{author}{\bibfnamefont{B.}~\bibnamefont{Nowak}},
  \bibinfo{author}{\bibfnamefont{D.}~\bibnamefont{Sexty}}, \bibnamefont{and}
  \bibinfo{author}{\bibfnamefont{T.}~\bibnamefont{Gasenzer}},
  \bibinfo{journal}{Phys. Rev. B} \textbf{\bibinfo{volume}{84}},
  \bibinfo{pages}{020506(R)} (\bibinfo{year}{2011}), \eprint{1012.4437}.

\bibitem[{\citenamefont{Nowak et~al.}(2012)\citenamefont{Nowak, Schole, Sexty,
  and Gasenzer}}]{Nowak:2011sk}
\bibinfo{author}{\bibfnamefont{B.}~\bibnamefont{Nowak}},
  \bibinfo{author}{\bibfnamefont{J.}~\bibnamefont{Schole}},
  \bibinfo{author}{\bibfnamefont{D.}~\bibnamefont{Sexty}}, \bibnamefont{and}
  \bibinfo{author}{\bibfnamefont{T.}~\bibnamefont{Gasenzer}},
  \bibinfo{journal}{Phys. Rev. A} \textbf{\bibinfo{volume}{85}},
  \bibinfo{pages}{043627} (\bibinfo{year}{2012}), \eprint{1111.6127}.

\bibitem[{\citenamefont{Gasenzer et~al.}(2012)\citenamefont{Gasenzer, Nowak,
  and Sexty}}]{Gasenzer:2011by}
\bibinfo{author}{\bibfnamefont{T.}~\bibnamefont{Gasenzer}},
  \bibinfo{author}{\bibfnamefont{B.}~\bibnamefont{Nowak}}, \bibnamefont{and}
  \bibinfo{author}{\bibfnamefont{D.}~\bibnamefont{Sexty}},
  \bibinfo{journal}{Phys. Lett.} \textbf{\bibinfo{volume}{B710}},
  \bibinfo{pages}{500} (\bibinfo{year}{2012}), \eprint{1108.0541}.

\bibitem[{\citenamefont{Schole et~al.}(2012)\citenamefont{Schole, Nowak, and
  Gasenzer}}]{Schole:2012kt}
\bibinfo{author}{\bibfnamefont{J.}~\bibnamefont{Schole}},
  \bibinfo{author}{\bibfnamefont{B.}~\bibnamefont{Nowak}}, \bibnamefont{and}
  \bibinfo{author}{\bibfnamefont{T.}~\bibnamefont{Gasenzer}},
  \bibinfo{journal}{Phys. Rev. A} \textbf{\bibinfo{volume}{86}},
  \bibinfo{pages}{013624} (\bibinfo{year}{2012}), \eprint{1204.2487}.

\bibitem[{\citenamefont{Karl et~al.}(2013)\citenamefont{Karl, Nowak, and
  Gasenzer}}]{Karl:2013mn}
\bibinfo{author}{\bibfnamefont{M.}~\bibnamefont{Karl}},
  \bibinfo{author}{\bibfnamefont{B.}~\bibnamefont{Nowak}}, \bibnamefont{and}
  \bibinfo{author}{\bibfnamefont{T.}~\bibnamefont{Gasenzer}}
  (\bibinfo{year}{2013}), \eprint{1302.1122}.

\bibitem[{\citenamefont{Berges et~al.}(2008)\citenamefont{Berges, Rothkopf, and
  Schmidt}}]{Berges:2008wm}
\bibinfo{author}{\bibfnamefont{J.}~\bibnamefont{Berges}},
  \bibinfo{author}{\bibfnamefont{A.}~\bibnamefont{Rothkopf}}, \bibnamefont{and}
  \bibinfo{author}{\bibfnamefont{J.}~\bibnamefont{Schmidt}},
  \bibinfo{journal}{Phys. Rev. Lett.} \textbf{\bibinfo{volume}{101}},
  \bibinfo{pages}{041603} (\bibinfo{year}{2008}), \eprint{0803.0131}.

\bibitem[{\citenamefont{Gasenzer and Pawlowski}(2008)}]{Gasenzer:2008zz}
\bibinfo{author}{\bibfnamefont{T.}~\bibnamefont{Gasenzer}} \bibnamefont{and}
  \bibinfo{author}{\bibfnamefont{J.~M.} \bibnamefont{Pawlowski}},
  \bibinfo{journal}{Phys. Lett.} \textbf{\bibinfo{volume}{B670}},
  \bibinfo{pages}{135} (\bibinfo{year}{2008}).

\bibitem[{\citenamefont{Scheppach et~al.}(2010)\citenamefont{Scheppach, Berges,
  and Gasenzer}}]{Scheppach:2009wu}
\bibinfo{author}{\bibfnamefont{C.}~\bibnamefont{Scheppach}},
  \bibinfo{author}{\bibfnamefont{J.}~\bibnamefont{Berges}}, \bibnamefont{and}
  \bibinfo{author}{\bibfnamefont{T.}~\bibnamefont{Gasenzer}},
  \bibinfo{journal}{Phys. Rev. A} \textbf{\bibinfo{volume}{81}},
  \bibinfo{pages}{033611} (\bibinfo{year}{2010}).

\bibitem[{\citenamefont{Schmidt et~al.}(2012)\citenamefont{Schmidt, Erne,
  Nowak, Sexty, and Gasenzer}}]{Schmidt:2012kw}
\bibinfo{author}{\bibfnamefont{M.}~\bibnamefont{Schmidt}},
  \bibinfo{author}{\bibfnamefont{S.}~\bibnamefont{Erne}},
  \bibinfo{author}{\bibfnamefont{B.}~\bibnamefont{Nowak}},
  \bibinfo{author}{\bibfnamefont{D.}~\bibnamefont{Sexty}}, \bibnamefont{and}
  \bibinfo{author}{\bibfnamefont{T.}~\bibnamefont{Gasenzer}},
  \bibinfo{journal}{New J. Phys.} \textbf{\bibinfo{volume}{14}},
  \bibinfo{pages}{075005} (\bibinfo{year}{2012}), \eprint{1203.3651}.

\bibitem[{\citenamefont{Micha and Tkachev}(2004)}]{Micha:2004bv}
\bibinfo{author}{\bibfnamefont{R.}~\bibnamefont{Micha}} \bibnamefont{and}
  \bibinfo{author}{\bibfnamefont{I.~I.} \bibnamefont{Tkachev}},
  \bibinfo{journal}{Phys. Rev. D} \textbf{\bibinfo{volume}{70}},
  \bibinfo{pages}{043538} (\bibinfo{year}{2004}), \eprint{hep-ph/0403101}.

\bibitem[{\citenamefont{Berges and Hoffmeister}(2009)}]{Berges:2008sr}
\bibinfo{author}{\bibfnamefont{J.}~\bibnamefont{Berges}} \bibnamefont{and}
  \bibinfo{author}{\bibfnamefont{G.}~\bibnamefont{Hoffmeister}},
  \bibinfo{journal}{Nucl. Phys.} \textbf{\bibinfo{volume}{B813}},
  \bibinfo{pages}{383} (\bibinfo{year}{2009}), \eprint{0809.5208}.

\bibitem[{\citenamefont{Berges and Sexty}(2011)}]{Berges:2010ez}
\bibinfo{author}{\bibfnamefont{J.}~\bibnamefont{Berges}} \bibnamefont{and}
  \bibinfo{author}{\bibfnamefont{D.}~\bibnamefont{Sexty}},
  \bibinfo{journal}{Phys. Rev. D} \textbf{\bibinfo{volume}{83}},
  \bibinfo{pages}{085004} (\bibinfo{year}{2011}), \eprint{1012.5944}.

\bibitem[{\citenamefont{Greensite}(2011)}]{Greensite:2011zz}
\bibinfo{author}{\bibfnamefont{J.}~\bibnamefont{Greensite}},
  \bibinfo{journal}{Lect. Notes Phys.} \textbf{\bibinfo{volume}{821}},
  \bibinfo{pages}{1} (\bibinfo{year}{2011}).

\bibitem[{\citenamefont{Ford et~al.}(1998)\citenamefont{Ford, Mitreuter, Tok,
  Wipf, and Pawlowski}}]{Ford:1998bt}
\bibinfo{author}{\bibfnamefont{C.}~\bibnamefont{Ford}},
  \bibinfo{author}{\bibfnamefont{U.}~\bibnamefont{Mitreuter}},
  \bibinfo{author}{\bibfnamefont{T.}~\bibnamefont{Tok}},
  \bibinfo{author}{\bibfnamefont{A.}~\bibnamefont{Wipf}}, \bibnamefont{and}
  \bibinfo{author}{\bibfnamefont{J.}~\bibnamefont{Pawlowski}},
  \bibinfo{journal}{Annals Phys.} \textbf{\bibinfo{volume}{269}},
  \bibinfo{pages}{26} (\bibinfo{year}{1998}), \eprint{hep-th/ 9802191}.

\bibitem[{\citenamefont{Braun et~al.}(2010)\citenamefont{Braun, Gies, and
  Pawlowski}}]{Braun:2007bx}
\bibinfo{author}{\bibfnamefont{J.}~\bibnamefont{Braun}},
  \bibinfo{author}{\bibfnamefont{H.}~\bibnamefont{Gies}}, \bibnamefont{and}
  \bibinfo{author}{\bibfnamefont{J.~M.} \bibnamefont{Pawlowski}},
  \bibinfo{journal}{Phys. Lett.} \textbf{\bibinfo{volume}{B684}},
  \bibinfo{pages}{262} (\bibinfo{year}{2010}), \eprint{0708.2413}.

\bibitem[{\citenamefont{Marhauser and Pawlowski}(2008)}]{Marhauser:2008fz}
\bibinfo{author}{\bibfnamefont{F.}~\bibnamefont{Marhauser}} \bibnamefont{and}
  \bibinfo{author}{\bibfnamefont{J.~M.} \bibnamefont{Pawlowski}}
  (\bibinfo{year}{2008}), \eprint{0812.1144}.

\bibitem[{\citenamefont{Gross et~al.}(1981)\citenamefont{Gross, Pisarski, and
  Yaffe}}]{Gross:1980br}
\bibinfo{author}{\bibfnamefont{D.~J.} \bibnamefont{Gross}},
  \bibinfo{author}{\bibfnamefont{R.~D.} \bibnamefont{Pisarski}},
  \bibnamefont{and} \bibinfo{author}{\bibfnamefont{L.~G.} \bibnamefont{Yaffe}},
  \bibinfo{journal}{Rev. Mod. Phys.} \textbf{\bibinfo{volume}{53}},
  \bibinfo{pages}{43} (\bibinfo{year}{1981}).

\bibitem[{\citenamefont{Weiss}(1981)}]{Weiss:1980rj}
\bibinfo{author}{\bibfnamefont{N.}~\bibnamefont{Weiss}},
  \bibinfo{journal}{Phys. Rev. D} \textbf{\bibinfo{volume}{24}},
  \bibinfo{pages}{475} (\bibinfo{year}{1981}).

\bibitem[{\citenamefont{Fister and Pawlowski}(2013)}]{Fister:2013bh}
\bibinfo{author}{\bibfnamefont{L.}~\bibnamefont{Fister}} \bibnamefont{and}
  \bibinfo{author}{\bibfnamefont{J.~M.} \bibnamefont{Pawlowski}}
  (\bibinfo{year}{2013}), \eprint{1301.4163}.

\bibitem[{\citenamefont{Diakonov et~al.}(2012)\citenamefont{Diakonov,
  Gattringer, and Schadler}}]{Diakonov:2012dx}
\bibinfo{author}{\bibfnamefont{D.}~\bibnamefont{Diakonov}},
  \bibinfo{author}{\bibfnamefont{C.}~\bibnamefont{Gattringer}},
  \bibnamefont{and} \bibinfo{author}{\bibfnamefont{H.-P.}
  \bibnamefont{Schadler}}, \bibinfo{journal}{JHEP}
  \textbf{\bibinfo{volume}{1208}}, \bibinfo{pages}{128} (\bibinfo{year}{2012}),
  \eprint{1205.4768}.

\bibitem[{\citenamefont{Greensite}(2012)}]{Greensite:2012dy}
\bibinfo{author}{\bibfnamefont{J.}~\bibnamefont{Greensite}},
  \bibinfo{journal}{Phys. Rev. D} \textbf{\bibinfo{volume}{86}},
  \bibinfo{pages}{114507} (\bibinfo{year}{2012}), \eprint{1209.5697}.

\bibitem[{\citenamefont{Langfeld and Pawlowski}(2013)}]{Langfeld:2013xbf}
\bibinfo{author}{\bibfnamefont{K.}~\bibnamefont{Langfeld}} \bibnamefont{and}
  \bibinfo{author}{\bibfnamefont{J.~M.} \bibnamefont{Pawlowski}}
  (\bibinfo{year}{2013}), \eprint{1307.0455}.

\bibitem[{\citenamefont{Felder et~al.}(2001)\citenamefont{Felder,
  Garcia-Bellido, Greene, Kofman, Linde et~al.}}]{Felder:2000hj}
\bibinfo{author}{\bibfnamefont{G.~N.} \bibnamefont{Felder}},
  \bibinfo{author}{\bibfnamefont{J.}~\bibnamefont{Garcia-Bellido}},
  \bibinfo{author}{\bibfnamefont{P.~B.} \bibnamefont{Greene}},
  \bibinfo{author}{\bibfnamefont{L.}~\bibnamefont{Kofman}},
  \bibinfo{author}{\bibfnamefont{A.~D.} \bibnamefont{Linde}},
  \bibnamefont{et~al.}, \bibinfo{journal}{Phys. Rev. Lett.}
  \textbf{\bibinfo{volume}{87}}, \bibinfo{pages}{011601}
  (\bibinfo{year}{2001}), \eprint{hep-ph/0012142}.

\bibitem[{\citenamefont{Asaka et~al.}(2001)\citenamefont{Asaka, Buchm{\"u}ller,
  and Covi}}]{Asaka:2001ez}
\bibinfo{author}{\bibfnamefont{T.}~\bibnamefont{Asaka}},
  \bibinfo{author}{\bibfnamefont{W.}~\bibnamefont{Buchm{\"u}ller}},
  \bibnamefont{and} \bibinfo{author}{\bibfnamefont{L.}~\bibnamefont{Covi}},
  \bibinfo{journal}{Phys. Lett.} \textbf{\bibinfo{volume}{B510}},
  \bibinfo{pages}{271} (\bibinfo{year}{2001}), \eprint{hep-ph/0104037}.

\bibitem[{\citenamefont{Sexty and Patkos}(2005{\natexlab{a}})}]{Sexty:2005pz}
\bibinfo{author}{\bibfnamefont{D.}~\bibnamefont{Sexty}} \bibnamefont{and}
  \bibinfo{author}{\bibfnamefont{A.}~\bibnamefont{Patkos}},
  \bibinfo{journal}{JHEP} \textbf{\bibinfo{volume}{0510}}, \bibinfo{pages}{054}
  (\bibinfo{year}{2005}{\natexlab{a}}), \eprint{hep-lat/0508014}.

\bibitem[{\citenamefont{{Derrick}}(1964)}]{Derrick1964a}
\bibinfo{author}{\bibfnamefont{G.~H.} \bibnamefont{{Derrick}}},
  \bibinfo{journal}{J. Math. Phys.} \textbf{\bibinfo{volume}{5}},
  \bibinfo{pages}{1252} (\bibinfo{year}{1964}).

\bibitem[{\citenamefont{Abrikosov}(1957)}]{Abrikosov1957a}
\bibinfo{author}{\bibfnamefont{A.}~\bibnamefont{Abrikosov}},
  \bibinfo{journal}{J. Phys. Chem. Solids} \textbf{\bibinfo{volume}{2}},
  \bibinfo{pages}{199} (\bibinfo{year}{1957}), ISSN \bibinfo{issn}{0022-3697}.

\bibitem[{\citenamefont{Nielsen and Olesen}(1973)}]{Nielsen:1973cs}
\bibinfo{author}{\bibfnamefont{H.~B.} \bibnamefont{Nielsen}} \bibnamefont{and}
  \bibinfo{author}{\bibfnamefont{P.}~\bibnamefont{Olesen}},
  \bibinfo{journal}{Nucl. Phys.} \textbf{\bibinfo{volume}{B61}},
  \bibinfo{pages}{45} (\bibinfo{year}{1973}).

\bibitem[{vid()}]{videos}
\bibinfo{note}{{\url{http://www.thphys.uni-heidelberg.de/~smp/videos/AbelianHiggs.html}}}.

\bibitem[{\citenamefont{Sexty and Patkos}(2005{\natexlab{b}})}]{Sexty:2005zm}
\bibinfo{author}{\bibfnamefont{D.}~\bibnamefont{Sexty}} \bibnamefont{and}
  \bibinfo{author}{\bibfnamefont{A.}~\bibnamefont{Patkos}},
  \bibinfo{journal}{Phys. Rev. D} \textbf{\bibinfo{volume}{71}},
  \bibinfo{pages}{025020} (\bibinfo{year}{2005}{\natexlab{b}}),
  \eprint{hep-ph/0501213}.

\end{thebibliography}

\end{document}